\documentclass[journal=jctcce,manuscript=article]{achemso}
\setkeys{acs}{articletitle = true}
\usepackage[T1]{fontenc}
\usepackage[utf8]{inputenc}
\usepackage{graphicx}
\usepackage{amsmath}
\usepackage{color}
\usepackage{comment}
\usepackage{hyperref}
\usepackage{multirow}
\usepackage{subfig}

\newcommand{\bC}{\mbox{\protect\boldmath $C$}}

\newcommand{\bD}{\mbox{\protect\boldmath $D$}}
\newcommand{\bB}{\mbox{\protect\boldmath $B$}}

\newcommand{\ket}{\rangle}
\newcommand{\bra}{\langle}

\def\AmS{{\the\textfont2 A}\kern-.1667em\lower.5ex\hbox
     {\the\textfont2 M}\kern-.125em{\the\textfont2 S}}

\def\AW{Addison\kern.1em-\penalty0\hskip0pt Wesley}
\def\BibTeX{{\rm B\kern-.05em{\smc i\kern-.025emb}\kern-.08em\TeX}}
\title{Hilbert space multireference coupled clusters tailored by matrix product states}

\author{Ond\v{r}ej Demel}
\affiliation{J. Heyrovsk\'{y} Institute of Physical Chemistry, Academy of Sciences of the Czech \mbox{Republic, v.v.i.}, Dolej\v{s}kova 3, 18223 Prague 8, Czech Republic}
\altaffiliation{Contributed equally to this work}

\author{Jan Brandejs}
\affiliation{J. Heyrovsk\'{y} Institute of Physical Chemistry, Academy of Sciences of the Czech \mbox{Republic, v.v.i.}, Dolej\v{s}kova 3, 18223 Prague 8, Czech Republic}
\alsoaffiliation{Faculty of Mathematics and Physics, Charles University, Prague, Czech Republic}
\altaffiliation{Contributed equally to this work}

\author{Jakub Lang}
\affiliation{J. Heyrovsk\'{y} Institute of Physical Chemistry, Academy of Sciences of the Czech \mbox{Republic, v.v.i.}, Dolej\v{s}kova 3, 18223 Prague 8, Czech Republic}
\alsoaffiliation{Faculty of Science, Charles University, Prague, Czech Republic}
\altaffiliation{Contributed equally to this work}

\author{Ji\v{r}\'{i} Brabec}
\affiliation{J. Heyrovsk\'{y} Institute of Physical Chemistry, Academy of Sciences of the Czech \mbox{Republic, v.v.i.}, Dolej\v{s}kova 3, 18223 Prague 8, Czech Republic}

\author{Libor Veis}
\affiliation{J. Heyrovsk\'{y} Institute of Physical Chemistry, Academy of Sciences of the Czech \mbox{Republic, v.v.i.}, Dolej\v{s}kova 3, 18223 Prague 8, Czech Republic}

\author{\"Ors Legeza} 
\affiliation{Strongly Correlated Systems ``Lend\"{u}let'' Research group, Wigner Research Centre for Physics, H-1525, Budapest, Hungary}
\alsoaffiliation{
Institute for Advanced Study,Technical University of Munich, Germany, Lichtenbergstrasse 2a, 85748 Garching, Germany
}

\author{Ji\v{r}\'{i} Pittner}
\email{jiri.pittner@jh-inst.cas.cz}
\affiliation{J. Heyrovsk\'{y} Institute of Physical Chemistry, Academy of Sciences of the Czech \mbox{Republic, v.v.i.}, Dolej\v{s}kova 3, 18223 Prague 8, Czech Republic}


\begin{document}

\begin{abstract}
In the last decade, the quantum chemical version of the density matrix renormalization group (DMRG) method has established itself as the method of choice for calculations of strongly correlated molecular systems. However, despite its favorable scaling, it is in practice not suitable for computations of dynamic correlation.
Several approaches to include that in post-DMRG methods exist; in our group we focused on the tailored-CC (TCC) approach, which employs
T$_1$ and T$_2$ amplitudes from the DMRG active space to externally correct the single-reference CC method.
This method works well in many situations, however, in exactly degenerate cases (with two or more determinants of equal weight), it
exhibits a bias towards the reference determinant representing the Fermi vacuum.
Although in some cases it is possible to use a compensation scheme to avoid this bias for energy differences,
as we did in a previous work on the singlet-triplet gap in the tetramethylenethane (TME) molecule, it is certainly a drawback.

In order to overcome the single-reference bias of the TCC method, we have developed a Hilbert-space multireference version of tailored CC,
which can treat several determinants on an equal footing. We have employed  a multireference analysis of the  DMRG wave function in the matrix product state form to get
the active amplitudes for each reference determinant and their constant contribution to the effective Hamiltonian.
We have implemented and compared the performance of three Hilbert-space MRCC variants - the state universal one, and the Brillouin-Wigner and Mukherjee's state specific ones.
We have assessed these approaches on the cyclobutadiene and tetramethylenethane (TME) molecules, which are both diradicals with exactly degenerate determinants
at a certain geometry.
Two DMRG active spaces have been selected based on orbital entropies, while the MRCC active space comprised the HOMO and LUMO orbitals
needed for description of the diradical.
We have also investigated the sensitivity of the results on orbital rotation of the HOMO-LUMO pair, as it is well known that
Hilbert-space MRCC methods are not invariant to such transformations.

\end{abstract}

\maketitle

\section{Introduction}

Quantum chemical computations became an indispensable part of the basic as well as applied research in chemistry during the last decades.
An effect of crucial importance for the accuracy of quantum chemical computations is the electron correlation.
As long as one Slater determinant dominates the wave function, only so-called dynamic correlation plays a role, and
the single-reference coupled cluster (CC) theory \cite{shavitt-bartlett}, originally  introduced to quantum chemistry by \v{C}\'{i}\v{z}ek \cite{cizek-original},
converges very fast with the excitation rank, already CCSD(T) \cite{ccsdtparent1} being accurate enough for many purposes.
However, the situation is dramatically different for strongly correlated systems,
where no single Slater determinant dominates the wave function and the usual single reference approaches become inaccurate or completely break down.
Since there are many chemically interesting compounds and reaction intermediates, which owing to their
quasidegenerate (and consequently multiconfigurational) nature belong to this class, the development of methods capable to treat such systems
is an important task.
Similarly, when one targets excited states, the  widely used methods like TDDFT or single-reference EOMCC become unreliable,
when the ground state is not dominated by a single Slater determinant.
An accurate description of such excited states is a challenging task, which requires new methodological progress and has many important applications \cite{barbattichemrev2018}.

One category of methods that aim at a balanced description of static and dynamic correlation are multireference coupled cluster methods, which generalize the CC exponential parameterization of the wave function.
Many approaches how to formulate the coupled cluster method for multireference problems have been suggested, which cannot be all listed here,
so we cite only several reviews \cite{bartlett-musial2007,ourspringer,bartlett_mrcc_2012,evangelista_perspective2018}.
Considerable effort has been put also into MRCC methods targeting excited states \cite{barbattichemrev2018}, however, even the advanced methods for excited states like icMRCC-LR \cite{icmrcclr} and MR-EOMCC \cite{mreom1} are limited to relatively  small active spaces \cite{barbattichemrev2018}.
Our group has focused on development of Hilbert space MRCC methods, mainly the state-specific ones --- Brillouin-Wigner MRCC (BWCC) and Mukherjee's MRCC (MkCC).
We have implemented the MkCC method with iteratively and perturbatively included triexcitations
\cite{mkcc_our1,mkcc-uncoupled-our,mkcc-uncoupled-t}.
In a collaboration with the group of J. Noga (Bratislava), we have developed explicitly correlated versions of the state-specific MRCC methods, which have the advantage of accelerated basis set convergence \cite{f12bw,mkccf12,mkccptf12}.
In order to compute large molecules and exploit modern supercomputer architectures, in collaboration with K. Kowalski we have developed and applied massively parallelized implementations  of BWCC and MkCC methods
\cite{bwpar,s_karolem_2012d,s_karolem_2012e,mkcc-parallel,s_karolem_2012c}
We have developed USS-based corrections to BWCC in a posteriori and iterative forms
\cite{s_karolem_2012b,uss_iterative}, which perform considerably better than the  previously used a posteriori BWCC correction \cite{bwcc-hubaccorrection}.
In a collaboration with F. Neese (Muelheim), who revived interest in the local pair natural orbital (LPNO) approaches and developed LPNO-based CC method \cite{local_cc_neese}.
We have generalized his ORCA  implementation to the multireference MkCC and BWCC methods at singles and doubles level \cite{ourlpno1}
and later developed also the near-linearly scaling DLPNO-based implementation of Hilbert-space MRCCSD \cite{dlpno-mkcc1} and MRCCSD(T) \cite{dlpno-mkcc2}.

Externally corrected coupled cluster methods, based on the single reference CC ansatz, represent another possible approach to strongly correlated systems.
They extract information about the most important higher excitations or active space mono- and biexcitations from an external calculation by a different method like CASSCF, MRCI, or DMRG \cite{paldus-externalcorr,paldus-externalcorr2,paldus-externalcorr-new1,paldus-externalcorr-new2,ec-1,PPacpq,ec-2a,tcc,cyclobut-tailored-2011,melnichuk-2012,melnichuk-2014,ourtcc2016,ourtcc2016erratum,ourtmebenchmark,lpnotcc2019,dlpnotcc2020,reltcc2020}.
Recently, Piecuch et al. combined the active space and method-of-moments based CC methods \cite{mmcc-first,bauman2017} with
input of a subset of higher amplitudes from the stochastic FCIQMC method by Alavi et al. \cite{booth_2009, cleland_2010, petruzielo_2012, overy_2015}, resulting in an externally corrected CC approach \cite{Piecuch-ccmc2017,Piecuch-ccmc2018,piecuch2021mccc}, which has been extended to excited states \cite{Piecuch-ccmc2019}.

DMRG is a very powerful approach suitable for treatment of strongly correlated systems, which has originally been developed by S.R. White in
solid state physics \cite{White-1992a,White-1992b,White-1993} and aimed mainly on treatment of one-dimensional model systems (chain of sites with interactions between neighbors).
The success of DMRG in this field motivated its application to quantum chemical problems\cite{Reiher-dmrg2020}, where
it has outperformed traditional CASSCF  for large active spaces.
In spite of the algorithmic progress in DMRG, it is computationally prohibitive to brute-force DMRG to treat the dynamic correlation
by including all virtual orbitals into the active space.
Since the dynamic correlation in general has a very significant chemical impact, development of ``post-DMRG''
methods aimed at describing this effect is an important topic and still a major challenge.
In Ref. \cite{Reiher-dmrg2020} an overview of various post-DMRG approaches is given.
Several competing approaches have been proposed: methods based on second order perturbation theory \cite{Kurashige-2011,Sharma-2014b,dmrg-nevpt2,dmrg-nevpt2b,dmrg-nevpt2c}, internally contracted MRCI \cite{Saitow-2013}, random phase approximation (RPA) \cite{Wouters-2013},
the canonical transformation method \cite{ctchan1,ctchan2,strongct,neuscamman_2010_irpc},
the adiabatic connection approach \cite{veis_adiabatic2021,pernal_adiabatic2022,pernal_adiabatic2023},
as well as the DMRG-tailored CC approach pursued in our group \cite{ourtcc2016,ourtcc2016erratum,ourtmebenchmark,lpnotcc2019} and recently also by others  \cite{Reiher-tcc2020,Boguslawski2021tcc}.
The TCC method did yield promising results, also for large systems employing the DLPNO-based implementation\cite{dlpnotcc2020}, and for molecules containing heavy atoms employing the
relativistic four-component implementation \cite{reltcc2020}.

However, TCC has also its disadvantages, probably the largest one being its single-reference bias, which particularly manifests itself in situations
when two (or more) determinants have almost degenerate weights.
In some situations of this kind it is possible to devise an error-compensation
scheme for energy difference quantities like e.g. the singlet-triplet gap, but
in general it seems advantageous to develop a multireference version of the TCC method, which is the topic of this work.

We first present the multireference DMRG-tailored CC based on the Hilbert-space MRCC approach and then also discuss the use of single-reference TCC in multiconfigurational situations, in particular in the case of biradicals.
Numerical assessment of the newly developed methods
on two interesting biradical molecules --- cyclobutadiene and tetramethyleneethane --- is then performed.

\section{Methods}

Before describing the newly developed multireference DMRG-tailored CC method
we will give a brief overview of the single reference DMRG-tailored CC and
Hilbert-space MRCC methods, which are its two essential ingredients.
Finally, we add some notes on using the single-reference tailored CC method in low spin open shell cases,
explaining the error compensation scheme for the singlet-triplet energy gap.

\subsection{The DMRG method and MPS wave function}

When applied to quantum chemical Hamiltonians, the DMRG method\cite{schollwock_2005,schollwock_2011,Legeza-2008,marti_2010,chan_review,wouters_review,legeza_review,yanai_review} runs on a ``lattice'' of active molecular orbitals,
where each orbital (``site'') has a 4-dimensional state space (0, $\alpha$-electron, $\beta$-electron, electron pair).
As shown by Rommer and \"Ostlund\cite{Ostlund-1995}, the DMRG sweep algorithm  corresponds to energy minimization with a wave function ansatz in
the matrix product state (MPS) form.
If there were no restrictions on the dimension of the matrices involved (bond dimension), DMRG would eventually converge to the FCI energy (at an exponential cost).

A practical version of DMRG for quantum chemistry is the two-site algorithm, which, in contrast to the one-site approach,
is less prone to get stuck in local minimum \cite{schollwock_2005}. It provides the wave function in the two-site MPS form \cite{schollwock_2011}
\begin{eqnarray}
  \label{mps_form}
  | \Psi_\text{MPS} \rangle = \sum_{\{\alpha\}} \mathbf{A}^{\alpha_1} \mathbf{A}^{\alpha_2} \cdots \mathbf{W}^{\alpha_i \alpha_{i+1}} \cdots \mathbf{A}^{\alpha_n}| \alpha_1 \alpha_2 \cdots \alpha_n \rangle, \nonumber \\
\end{eqnarray}
where $\alpha_i \in \{ | 0 \rangle, | \downarrow \rangle, | \uparrow \rangle, | \downarrow \uparrow \rangle \}$ and
for a given pair of adjacent indices $[i, (i+1)]$, $\mathbf{W}$ is a four index tensor, which corresponds to the eigenfunction of the electronic Hamiltonian expanded in the tensor product space of four
tensor spaces defined on an ordered orbital chain, so-called \textit{left block} ($M_l$ dimensional tensor space)  , \textit{left site} (four dimensional tensor space of $i^{\text{th}}$ orbital), \textit{right site} (four dimensional tensor space of $(i+1)^{\text{th}}$ orbital), and \textit{right block} ($M_r$ dimensional tensor space).
The MPS matrices $\mathbf{A}$ are obtained by successive application of the singular value decomposition (SVD) with truncation on $\mathbf{W}$'s and iterative optimization by
going through the ordered orbital chain from \textit{left} to \textit{right} and then sweeping back and forth.
When employing the two-site MPS wave function (Eq. \ref{mps_form}) for the purposes of the TCCSD method,
the CI expansion coefficients $c_i^a$ and $c_{ij}^{ab}$ for $a,b,i,j \in \text{CAS}$, as well as coefficients of higher excitations can be efficiently calculated by contractions of MPS matrices \cite{moritz_2007, boguslawski_2011}.


\subsection{The single-reference DMRG-TCC method}

The tailored CC (TCC) approach pioneered by Kinoshita et al. \cite{kinoshita_2005} belongs to the class of so called externally corrected CC methods.
The TCC wave function assumes the split-amplitude exponential form introduced previously by Piecuch et al. \cite{piecuch_1993, piecuch_1994}
\begin{equation}
  \label{eq:TCC}
  | \Psi_\text{TCC} \rangle = e^{T} | \Phi_\text{0} \rangle =  e^{T_\text{ext}+T_\text{CAS}} | \Phi_\text{0} \rangle =  e^{T_\text{ext}} e^{T_\text{CAS}} | \Phi_\text{0} \rangle,
\end{equation}
i.e. the cluster operator is split up into its active space part ($T_\text{CAS}$) and the remaining external part ($T_\text{ext}$). Since $| \Phi_\text{0} \rangle$ is a single-determinant reference wave function, both of the aforementioned cluster operators mutually commute, which keeps the methodology very simple.
The $T_\text{CAS}$ amplitudes are supposed to be responsible for a proper description of the static correlation. They are computed from the MPS wave function, first obtaining the CI coefficients and then performing a ``cluster analysis'' using the well known relations between the linear CI and exponential CC expansions.
\begin{eqnarray}
        \label{eq:ci2cc}
        T^{(1)}_{\text{CAS}} & = & C^{(1)}, \\
        T^{(2)}_{\text{CAS}} & = & C^{(2)} - \frac{1}{2}[C^{(1)}]^2.
\end{eqnarray}
The $T_\text{CAS}$ amplitudes are kept frozen during the CC iterations, while the $T_\text{ext}$ part, which is responsible for a proper description of the dynamic correlation, is being optimized.

\subsection{A brief overview of the Hilbert-space MRCC methods}

The Hilbert-space MRCC methods\cite{bartlett-musial2007,ourspringer,bartlett_mrcc_2012}
are based on the Jeziorski-Monkhorst (JM) Ansatz\cite{jezmonk} 
\begin{equation}
\label{omegaansatz}
\Omega_\alpha = \sum_{\mu=1}^M  e^{T_\alpha(\mu)} |\Phi_\mu\ket\bra \Phi_\mu|
\end{equation}
for the wave operator $\Omega_\alpha$, which maps a model-space projection of the exact wave function of state $\alpha$ back to the exact wave function
\begin{equation}
\label{waveoperdef}
   \Psi_\alpha   =  \Omega_\alpha   \Psi_\alpha^P \equiv  P\Psi_\alpha   =  \sum_{\mu=1}^{M} c^\alpha_\mu |\Phi_\mu \rangle
\end{equation}
Here $M$ is the dimension of the model (reference) space spanned by the Slater determinants $|\Phi_\mu\rangle$ and $P$ is a projector on this space.
In most formulations one requires the wave operator to obey the intermediate normalization, which translates to the condition
\begin{equation}
\label{intermednorm}                                                                    P\Omega_\alpha P = P\Omega_\alpha  = P                   
\end{equation}
Using the wave operator (\ref{waveoperdef}) it is possible to introduce an effective Hamiltonian
\begin{equation}
\label{heff}
H^{\rm eff}_\alpha = P H \Omega_\alpha P
\end{equation}
This Hamiltonian operates on wave functions from the model space, but shares eigenvalue(s)
with the full Hamiltonian
\begin{equation}
\label{heffeig}                                                                              H^{\rm eff}_\alpha \Psi_\alpha^P = E_\alpha  \Psi_\alpha^P
\end{equation} 
If this holds for all eigenvalues, the resulting method is called state-universal,
otherwise it is called state-specific for state no. $\alpha$.
The state-universal MRCC method (SUMRCC) can be derived by inserting the JM Ansatz
into the Bloch equation 
\begin{equation} 
\label{rsbloch2}
H \Omega P = \Omega P H \Omega P \equiv  \Omega H^{\rm eff}
\end{equation}
which is derived from Schr\"{o}dinger equation but eliminates the explicit appearance of the energy.
On the other hand, the state-specific methods are derived directly from the Schr\"{o}dinger equation. Since they only target one state, their derivation leads to an underdetermined
set of equations, which has to be augmented by ``sufficiency conditions'' to yield equations
for the individual amplitudes in the JM Ansatz (\ref{omegaansatz}). Depending on the details of these conditions,
different state-specific methods can be obtained.

The resulting amplitude equations for the state-universal method\cite{jezmonk} read
\begin{equation}
\label{suequations2}
\bra \Phi_\vartheta| H_N(\mu) e^{T(\mu)} |\Phi_\mu \ket_C \equiv \bra \Phi_\vartheta| e^{-T(\mu)} H_N(\mu) e^{T(\mu)} |\Phi_\mu \ket =  \sum_{\nu \neq \mu} \bra \Phi_\vartheta| e^{-T(\mu)} e^{T(\nu)} |\Phi_\nu \ket H^{\rm eff}_{\nu\mu}
\end{equation}
where $\Phi_\vartheta$ is an excited determinant with respect to the $\Phi_\mu$ Fermi vacuum and the subscript C indicates that only connected diagrams are taken into account.
An alternative formulation derived by Kucharski and Bartlett \cite{kucharski-mrcc} exists,
which is equivalent to (\ref{suequations2}) if the model space is complete.

The simplest state-specific method is the Brillouin-Wigner one (BWCC) \cite{hubac-originalbwcc,hubac-neogrady,masik-diagrams,hubac-originalbwcc2,MasikSanMini,bwcc1}
where the amplitude equations for different reference are coupled via the energy
\begin{equation}
(E_\alpha-H^{\rm eff}_{\mu\mu}) \bra \Phi_\vartheta| e^{T(\mu)} |\Phi_\mu\ket  = \bra \Phi_\vartheta| H_N(\mu)  e^{T(\mu)} |\Phi_\mu\ket_{C+DC,L}
\label{bwequationsfinal}
\end{equation}
where the subscript $_{C+DC,L}$ indicates inclusion of both connected and disconnected but linked diagrams.
The disadvantage of the BWCC method is its lack of size-extensivity, which requires a correction\cite{bwcc-hubaccorrection,pittner-continuous}.
This issue has been addressed by Mukherjee et al. \cite{mukherjee}, who derived
a size-extensive state specific MRCC version (MkCC) with amplitude equations
\begin{equation} \label{mk-cl-eq}
\langle   \Phi_\vartheta(\mu)| e^{-T(\mu)} H e^{T(\mu)} | \Phi_\mu \rangle c_\mu^\alpha+
\sum_{\nu \ne \mu} H^{\rm eff}_{\mu \nu} c_\nu^\alpha \langle \Phi_\vartheta(\mu) |
e^{-T(\mu)} e^{T(\nu)} | \Phi_\mu \rangle = 0
\end{equation}
The last possibility was derived by Mahapatra and Chattopadhyay \cite{mkcc_variant_mahapatra,tmm-mahapatra} under the name H2E2 MRCC with amplitude equations
\begin{eqnarray} \label{h2e2final} 
\bra \Phi_\vartheta(\mu)|  e^{-T(\mu)} H e^{T(\mu)} |\Phi_\mu \rangle c^{\alpha}_\mu  &+&
 \sum_{\nu\;(\neq \mu)} \left[  
\bra \Phi_\vartheta(\mu)| e^{-T(\mu)} e^{T(\nu)} |\Phi_\mu \rangle H^{\rm eff}_{\mu \nu}c^{\alpha}_\nu
\right. - \\ 
\nonumber 
&-& \left.  \bra  \Phi_\vartheta(\mu)| e^{-T(\mu)}  e^{T(\nu)} |\Phi_\nu\ket  H^{\rm eff}_{\nu \mu}  c^{\alpha}_\mu
\right]  =0.
\end{eqnarray}
The relationship between different Hilbert-space MRCC methods has been studied by Kong  \cite{lkong_inv}.

\subsection{Multireference DMRG-tailored CC}

The TCC method is based on the single-reference CC ansatz and exhibits thus a bias if the wave function
is dominated by two or more determinants with a comparable weight.
To ameliorate this we propose to combine the tailored CC idea with the Jeziorski-Monkhorst MRCC ansatz
\begin{eqnarray}
\label{mrtccansatz}
|\Psi^\alpha\rangle &=& \sum_{\mu \in {\cal M}}^M C_\mu^\alpha |\Psi_\mu\rangle =  \sum_{\mu \in {\cal M}}^M C_\mu^\alpha e^{T(\mu)} |\Phi_\mu\rangle\\
T(\mu) &=& T_{\rm CAS}(\mu) + T_{\rm ext}(\mu)
\end{eqnarray}
where the orbital set used to build the $M$-dimensional Jeziorski-Monkhorst model space ${\cal M}$ is a subset of the (substantially) larger CAS
in which DMRG is performed and used as the external source of amplitudes.

In order to uniquely obtain $T_{\rm CAS}(\mu)$ amplitudes for all $\mu$ from the CASCI/MPS wave function
one has to consider $M$ states $\Psi_\alpha$ and perform a MR generalization of the CC analysis of a CI wave function \cite{paldus-ccanalysis}.
Then the coefficients $C_\mu^\alpha$ will form an invertible matrix $\bC = ||C_\mu^\alpha||$ and
\begin{equation}
\label{mrccanalysis1}
|\Psi_\mu\rangle \equiv e^{T(\mu)} |\Phi_\mu\rangle = \sum_{\alpha = 1}^M (\bC^{-1})_\mu^\alpha |\Psi^\alpha\rangle
\end{equation}
Expanding the MRCC wave function $|\Psi_\mu\rangle $ in determinants from the model space and its orthogonal complement yields
\begin{equation}
\label{mrccanalysis2}
|\Psi^\alpha\rangle =  \sum_{\mu \in {\cal M}} C_\mu^\alpha  |\Phi_\mu\rangle + \sum_{\vartheta \notin {\cal M}}  D_\vartheta^\alpha  |\Phi_\vartheta\rangle
\end{equation}
Multiplying the vector of $\{|\Psi^\alpha\rangle \}_{\alpha=1}^M$ by the inverse matrix $\bC^{-1}$ we get
\begin{equation}
\label{mrccanalysis3}
|\Psi_\mu\rangle =  |\Phi_\mu\rangle + \sum_{\vartheta \notin {\cal M}}  B_\vartheta^\mu   |\Phi_\vartheta\rangle
\end{equation}
where the coefficient matrix $\bB \equiv ||B_\vartheta^\mu || = \bC^{-1} \bD$.
The reference-specific wave function $|\Psi_\mu\rangle$ has a single-reference-like structure and can be analyzed in the usual way comparing the CI coefficients and products of amplitudes resulting from the exponential
\begin{eqnarray}
\label{analtcas1}
T_{\rm CAS,1}(\mu) &=& B_1(\mu) \\
\label{analtcas2}
T_{\rm CAS,2}(\mu) &=& B_2(\mu) - 1/2 T_{\rm CAS,1}(\mu)^2 \\
\label{analtcas3}
T_{\rm CAS,3}(\mu) &=& B_3(\mu) - T_{\rm CAS,1}(\mu) T_{\rm CAS,2}(\mu) -1/6 T_{\rm CAS,1}(\mu)^3 \\
\nonumber
&\ldots&
\end{eqnarray}

The DMRG calculation yields wave functions in the matrix product state (MPS) form, from which CI coefficients with a limited excitation
rank can be obtained efficiently by performing contractions of the virtual indices in MPS for selected values of physical indices.
In particular, the {\em efficient} MRCC  analysis takes advantage of the compressed MPS form of the DMRG
wave functions. All the targeted states indeed share the MPS matrices except for the two active sites (two-site DMRG). Moreover, these
two-site tensors are for the purposes of diagonalization internally stored in the form of vectors, which can be easily rotated with the
inverted matrix of reference coefficients in all targeted states. The CI coefficients and subsequently CC
amplitudes for all the references are then obtained analogously to the single reference case from these rotated MPS wave functions \cite{ourtcc2016,ourtcc2016erratum}.

From this point onward, all $T_{\rm CAS}(\mu)$ amplitudes are kept fixed in the tailored CC spirit, while $ T_{\rm ext}(\mu)$ are obtained iteratively
using one of the Hilbert-space MRCC schemes (Mukherjee's, Brillouin-Wigner, or state-universal).

However, in contrast to the single-reference TCC, there is a slight complication with the effective Hamiltonian
\begin{equation}
\label{heff_mrtcc}
H^{\rm eff}_{\nu\mu} = \langle \Phi_\nu | H_N(\mu) e^{T_{\rm CAS}(\mu) + T_{\rm ext}(\mu)} |\Phi_\mu\rangle_C
\end{equation}
where $H_N(\mu)$ is the Hamiltonian in normal order form with respect to $\Phi_\mu$ Fermi vacuum and the $C$ subscript indicates that only connected diagrams contribute (for complete model space).
Its diagonal elements, being diagrammatically analogous to the single-reference CC energy formula, depend explicitly only on $T_1(\mu)$ and $T_2(\mu)$, regardless of the truncation of $T$ and mutual excitation rank between $\Phi_\nu$ and $\Phi_\mu$ and are thus straightforward.
On the other hand, higher $T_{\rm CAS}(\mu)$ than biexcitations contribute in general to the off-diagonal elements, even if $T_{\rm ext}$ is truncated to SD, and we have to account for this contribution.
Otherwise, the MRTCC energy obtained by the diagonalization of $H^{\rm eff}$, would not recover the DMRG (CASCI) energy for $T_{\rm ext}(\mu)=0$, which
is a sensible consistency requirement that is fulfilled in the single-reference case.
Rather than performing the MRCC analysis for $T_{\rm CAS,3}(\mu)$, $T_{\rm CAS,4}(\mu)$, \ldots and computing explicitly their contribution to $H^{\rm eff}_{\nu\mu}$,
we use the fact that $T_{\rm CAS}(\mu)$ are fixed during the MRTCC iterations. They thus contribute to $H^{\rm eff}_{\nu\mu}$  by
a constant additive term, and in general also by  terms mixing $T_{\rm CAS}(\mu)$ and $T_{\rm ext}(\mu)$, which however vanish for $T_{\rm ext}(\mu)=0$.
The constant contribution of $T_{{\rm CAS},n}(\mu), \;\;(n>2)$ can be expressed as
\begin{equation}
\label{heffcorr}
H^{\rm eff, corr}_{\nu\mu} =  \langle \Phi_\nu | H_N(\mu) e^{T_{\rm CAS}(\mu)}|\Phi_\mu\rangle_C - \langle \Phi_\nu | H_N(\mu) e^{T_{\rm CAS,1}(\mu) + T_{\rm CAS,2}(\mu)}|\Phi_\mu\rangle_C
\end{equation}
where the second term can be easily obtained in the first MRTCCSD iteration when $T_{\rm ext}(\mu)=0$.
To obtain the first term from DMRG/CASCI calculation, not much additional effort is needed, since
\begin{equation}
\langle \Phi_\nu | H_N(\mu) e^{T_{\rm CAS}(\mu)}|\Phi_\mu\rangle_C  = \langle \Phi_\nu |H_N(\mu)| \Psi_\mu^{\rm CAS}\rangle_C
\end{equation}
and the wave function $|\Psi_\mu^{\rm CAS}\rangle \equiv  e^{T_{\rm CAS}(\mu)}|\Phi_\mu\rangle$ is needed for the MRCC analysis (\ref{mrccanalysis3}-\ref{analtcas3}) anyway.
The effective Hamiltonian is then computed during the MRTCCSD iterations as
\begin{equation}
\label{hefffinal}
H^{\rm eff}_{\nu\mu} = H^{\rm eff, corr}_{\nu\mu} +  \langle \Phi_\nu | H_N(\mu) e^{T_{\rm CAS,12}(\mu) + T_{\rm ext,12}(\mu)} |\Phi_\mu\rangle_C
\end{equation}
The correction (\ref{heffcorr}) lacks some terms combining $T_{\rm ext}(\mu)$ and $T_{\rm CAS}(\mu)$ which become nonzero during the MRTCC iterations; for example for mutually biexcited
$\Phi_\nu$ and $\Phi_\mu$, the term $\langle \Phi_\nu|H_{N,2} T_{\rm CAS,3}(\mu) T_{\rm ext,1}(\mu)|\Phi_\mu\rangle$ is missing.
Such terms will, however, not affect the limiting cases where MRTCC should reproduce CASCI or FCI energies, for $T_{\rm ext}=0$ or for the CAS space spanning all orbitals, respectively. We think that it is thus a justified approximation to neglect them.

\subsection{A note on the use of single-reference tailored CC in low-spin open shell cases}

The single-reference tailored CC method (as well as standard CC) exhibits a bias, energetically favoring systems where there is one dominant determinant used as its reference compared to systems where two (or more) determinants have an equal weight. As a consequence, it breaks the energy degeneracy of the individual components of a triplet state with different $M_S$. However, this feature
can be used to devise an error compensation scheme for calculations of energy differences.
In the simplest case of a diradical, which can be described by the CAS(2,2) space consisting of two electrons in the HOMO-LUMO pair ($\phi_H,\,\phi_L$), the singlet-triplet energy gap can be reliably computed as a difference of the energy of the singlet and the energy of the $M_S=0$ component of the triplet, since both states are then described ``equally poorly'' by the single-reference method. Use of  $M_S=1$ triplet energy might on the other hand yield even a wrong sign of the ST gap with triplet too low. \cite{ourtmebenchmark}

The $M_S=0$ triplet component is dominated by two ``low-spin'' Slater determinants
\begin{equation}
\label{tripletms0}
\Psi_{S=1,\,M_S=0} = \frac{1}{\sqrt 2} \left({\rm det}|\phi_H\alpha, \phi_L\beta| - {\rm det}  |\phi_H\beta, \phi_L\alpha|\right)
\end{equation}
Henceforth, we will abbreviate such a wavefunction as $(ab-ba)$ or $(\ldots 2ab - \ldots 2ba)$ if
the preceding inactive electrons should be emphasized.
It is possible to compute such a state using a standard CC code, which assumes either a closed shell or a high-spin open shell reference, as long as the code has a spin-unrestricted implementation
supporting different spatial orbitals for $\alpha$ and $\beta$ spins.
One can then exchange the $\beta$ spinorbitals of HOMO and LUMO, 
$\phi_H' = \phi_L$, $\phi_L' = \phi_H$
so that the wave function
(\ref{tripletms0}) will be transformed to 
\begin{equation}
\label{tripletms0tr}
\Psi_{S=1,\,M_S=0} =\frac{1}{\sqrt 2} \left( {\rm det}|\phi_H\alpha, \phi_H'\beta| - {\rm det}  |\phi_L'\beta, \phi_L\alpha|\right)
\end{equation}
abbreviated as $(20-02)$. Clearly one can then use the single reference method with the Fermi vacuum $20$, although the description of the state is poor if the two determinants are quasidegenerate.

Actually, this technique can be generalized for treatment of any wave function with $M_S \geq 0$.
One chooses the (low-spin) reference determinant and then performs separate permutations of both $\alpha$ and $\beta$ spinorbitals which bring the chosen determinant to the form $(\ldots 22a \ldots a 00\ldots)$. In the tailored CC approach, this can be sometimes used to target both ground and excited state, as exemplified below.

Since the single-reference CC and TCC methods are not invariant to rotations between HOMO and LUMO
and the Hilbert space MRCC methods are not invariant with respect to rotations among the active orbitals, it is interesting also to investigate the influence of such  rotations on the resulting energies. 
We have investigated the particularly interesting $\pi/4$ mixing angle, when
\begin{eqnarray}
\label{rotation45}
\phi_H &=& \frac{1}{\sqrt 2} ( \tilde{\phi}_H  +  \tilde{\phi}_L)\\
\nonumber
\phi_L &=& \frac{1}{\sqrt 2} (\tilde{\phi}_H  -  \tilde{\phi}_L)\\
\nonumber
\end{eqnarray}
Such an orbital rotation of course transforms also the many-electron reference wave function and one has to account for this in the computational setup. 
For example, the singlet wave function which in the original orbitals had the character $(20-02)$ 
\begin{equation}
\label{singlet}
\Psi_{S=0,\,M_S=0} = \frac{1}{\sqrt 2} \left({\rm det}|\phi_H\alpha, \phi_H\beta| - {\rm det}  |\phi_L\alpha, \phi_L\beta|\right)
\end{equation}
will after inserting the transformation (\ref{rotation45}) become $(ab+ba)$
\begin{equation}
\label{singlet2}
\Psi_{S=0,\,M_S=0} =\frac{1}{\sqrt 2} ( {\rm det}|\tilde{\phi}_H\alpha, \tilde{\phi}_L\beta| + {\rm det}  |\tilde{\phi}_H\beta, \tilde{\phi}_L\alpha| )
\end{equation}
This transformation can also be combined with the aforementioned swap of $\beta$ spinorbitals to
bring the transformed wave function into a form suitable for calculation with a code assuming
a high-spin reference determinant.

\begin{table}[ht]
\caption{\label{table_trafo} Wave function character of the lowest singlet and triplet states in a CAS(2,2) space subject to $\pi/4$ orbital mixing and swap of $\beta$ spinorbitals}
\begin{tabular}{lcccc}
\hline
Transformation &none &rotation&swap&rotation+swap\\
\hline
Singlet&$(20-02)$&$(ab+ba)$&$(ab-ba)$&$(20+02)$\\
Triplet&$(ab-ba)$&$(ab-ba)$&$(20-02)$&$(20-02)$\\
\hline
\end{tabular}
\end{table}

We have performed a systematic investigation of the influence of the orbital rotation on the
single- as well as multireference tailored CC energies. In Table~\ref{table_trafo} we list the
resulting characters of the wave functions of the lowest singlet and triplet states in the CAS(2,2)
space subject to the aforementioned transformations.
Notice, in particular, the interesting fact that when the $\pi/4$ rotation is combined with the
$\beta$ HOMO-LUMO spinorbital swap, both singlet and triplet state acquire a formally closed shell
character (with different orbitals for different spins, of course). This has the interesting side effect, that the single-reference tailored CC method with closed shell reference 20 can be employed to compute both states, because
the double excitation amplitude which corresponds to the other dominant determinant is frozen during the TCC calculation, and (due to the intermediate normalization) is approximately $+1$ or $-1$, respectively. Standard single-reference CC,
which iterates this amplitude, converged to the $M_S=0$ triplet component, although this behavior might depend on numerical details.

\section{Numerical results}

\subsection{Automerization of cyclobutadiene}

\begin{figure}[!ht]
  \includegraphics[width=0.45\textwidth]{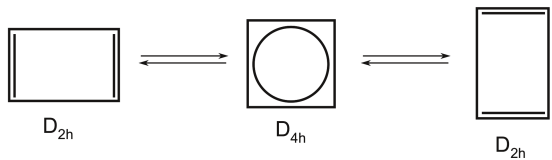}
  \caption{Automerization reaction of cyclobutadiene.}
  \label{fig_cb_reaction}
\end{figure}

  Cyclobutadiene is a fascinating molecule that over the years attracted a lot of attention of theoretical as well as experimental chemists.
  It undergoes an automerization reaction, see Figure \ref{fig_cb_reaction}, for which the carbon atom tunneling mechanism has been proposed \cite{michl_review, michl_jacs, arnold_1993}. It is also important for its anti-aromatic character and alleged aromatic character of its first excited triplet state, which makes it particularly interesting \cite{cyclobut-rectangular-orbitals,szalay_review}.
  The automerization reaction of cyclobutadiene is indeed a perfect benchmark for multireference methods\cite{balkova-cyclobut,cyclobut,cyclobut-tailored-2011,paldus-cyclobut-2009,cyclobut-barbatti} as the $D_{2h}$ ground state ($^1A_{1g}$) is a closed-shell system, but the $D_{4h}$ transition state is an open-shell system with degenerate orbitals, thus requiring a proper multireference treatment \cite{szalay_review}.
  The experimental values for the automerization barrier vary considerably within the range $1.6-12$ kcal/mol \cite{carpenter_1982}. We hope that our results will shed more light on this delicate property and eventually help to determine its value more accurately.
  The aforementioned degeneracy in case of the transition state is a perfect test bed for the multireference tailored CC methods.
  In what follows, we will abbreviate the spectroscopic labeling of electronic states in the following (and usual) way: 1$^{1}$A$_{\text{g}}$ $\rightarrow$ S$_0$, 1$^{3}$B$_{\text{1g}}$ $\rightarrow$ T$_1$, 1$^{1}$B$_{\text{1g}}$ $\rightarrow$ S$_1$, 2$^{1}$A$_{\text{g}}$ $\rightarrow$ S$_2$.

\subsubsection{Computational details}

  The geometries were taken from \cite{em_2006} and correspond to the MR AQCC/cc-pVTZ methodology with state averaged CASSCF(4,4) orbitals. In particular, the rectangular ground state geometry with $r_{\text{C=C}} = 1.349$ \AA~and $r_{\text{C-C}} = 1.562$ \AA~was linearly interpolated to the square transition state geometry with $r_{\text{C-C}} = 1.447$ \AA~(by means of $\lambda$ parameter).
   We have employed the CASPT2(4,4)/cc-pVTZ natural orbitals state averaged over all the states in study (S$_0$, T$_1$, S$_1$, S$_2$). The active space contained two $\pi$ and two $\pi^{*}$ orbitals.
  Based on the occupation numbers of CASPT2(4,4) natural orbitals, we have selected two DMRG active spaces: CAS(20,14) (occupation numbers larger than 0.015) and CAS(20,19) (occupation numbers larger than 0.01).
  For CASPT2 calculations, we have employed the MOLPRO package \cite{molpro}, DMRG calculations were carried out with Budapest QCDMRG code \cite{budapest_qcdmrg} and TCC and MRTCC calculations were performed with our implementation in ORCA \cite{orca}.

\subsubsection{Results and Discussion}

\begin{figure}[!ht]
  \subfloat[CAS(20,14)\label{subfig_cyclobut_14}]{%
    \includegraphics[width=0.48\textwidth]{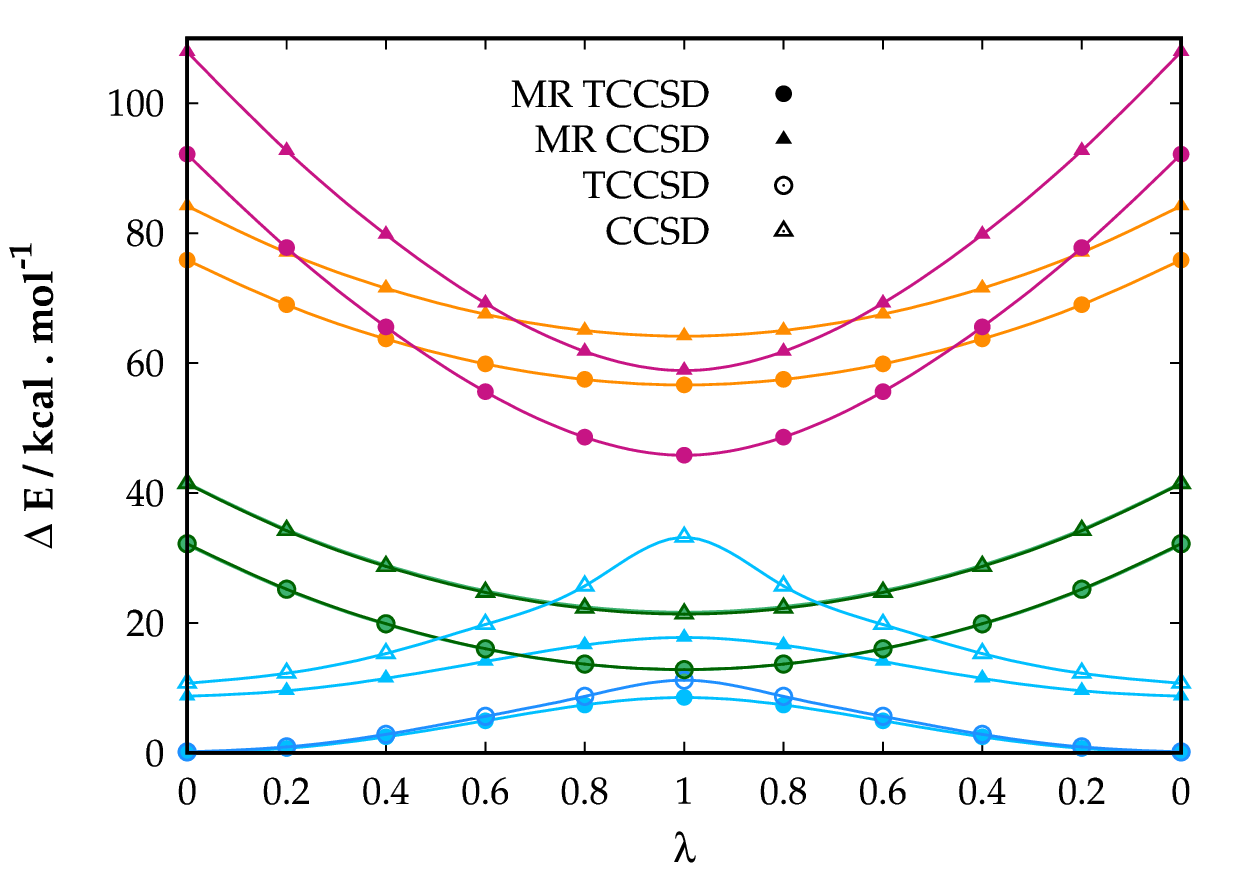}
  }
  \hfill
  \subfloat[CAS(20,19)\label{subfig_cyclobut_19}]{%
    \includegraphics[width=0.48\textwidth]{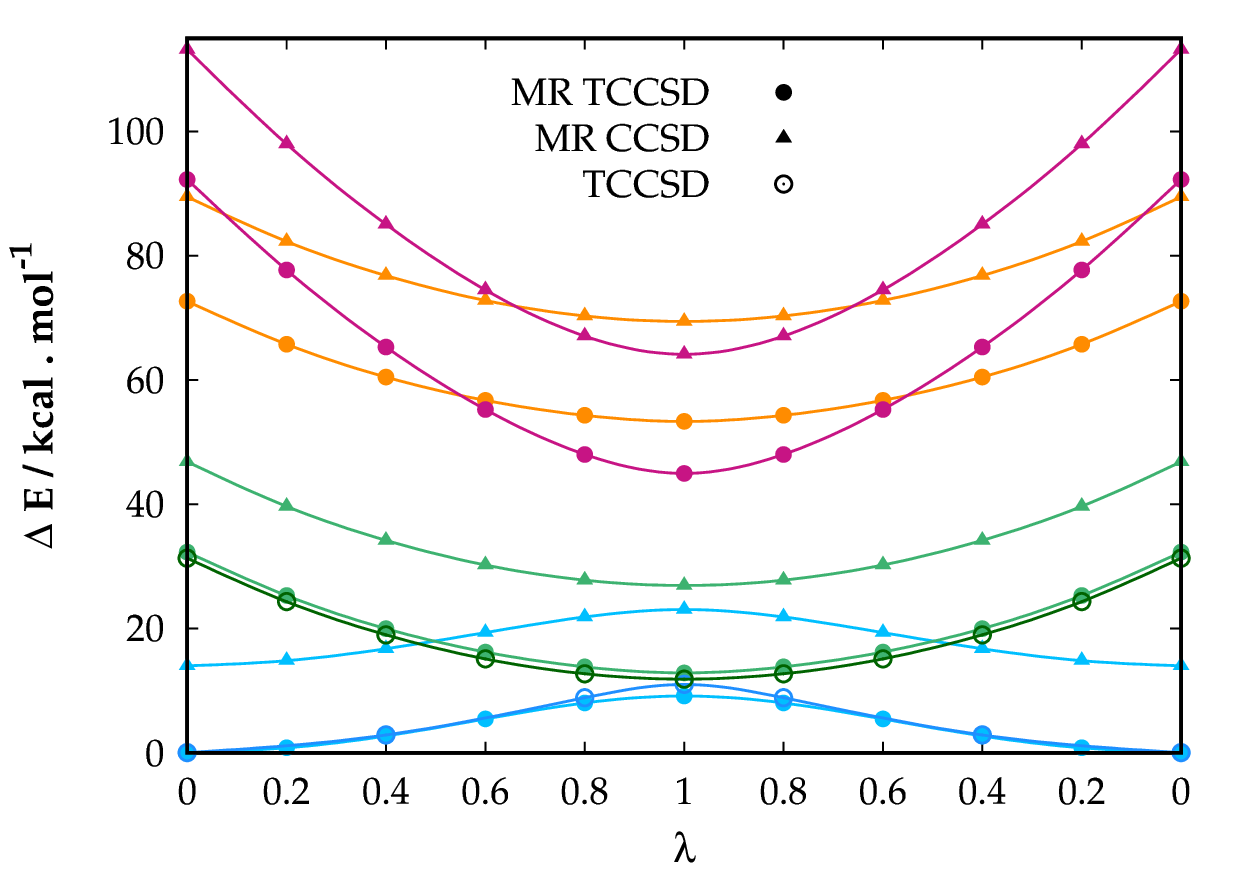}
  }
  \caption{The potential energy surfaces of the four lowest lying states: S$_0$ (blue), T$_1$ (green), S$_1$ (orange), and S$_2$ (purple) along the automerization path. The left subfigure corresponds to the DMRG active spaces containing 14 orbitals, whereas the right to the active spaces containing 19 orbitals.}
  \label{fig_cyclobut}
\end{figure}

\begin{table}[!ht]
  \begin{tabular}{r r}
  \hline
  \hline
   & $\Delta E_{\text{el}}$ \\
  \hline 
  MR AQCC \cite{em_2006} & 8.4 \\
  CCSD & 22.46 \\
  MR CCSD &  9.06 \\ 
  TCCSD(20,14) & 11.04 \\ 
  MR TCCSD(20,14) & 8.52 \\ 
  TCCSD(20,19) & 10.99 \\
  MR TCCSD(20,19) & 9.13 \\ 
  \hline
  \hline
  \end{tabular}
  \caption{The MR CCSD, MR TCCSD, and MR AQCC \cite{em_2006} automerization energy barriers in kcal mol$^{-1}$. All the results correspond to the cc-pVTZ basis \cite{dunning_basis}.}
  \label{tab_cyclobut_barriers}
\end{table}

\begin{table}[!ht]
  \begin{tabular}{r c c c}
  \hline
  \hline
   & S$_0$ $\rightarrow$ T$_1$ & S$_0$ $\rightarrow$ S$_1$ & S$_0$ $\rightarrow$ S$_2$ \\
  \hline 
  MR AQCC \cite{em_2006} & 5.5 & 31.5 & 43.3 \\
  MR CCSD & 3.8 & 48.2 & 46.4 \\ 
  MR TCCSD(20,14) & 4.3 & 37.5 & 48.1 \\ 
  MR TCCSD(20,19) & 3.7 & 36.0 & 44.1 \\ 
  \hline
  \hline
  \end{tabular}
  \caption{The MR CCSD, MR TCCSD, and MR AQCC \cite{em_2006} energy gaps in kcal mol$^{-1}$ between the transition state and the lowest three excited states optimized at MR AQCC level \cite{em_2006}. All the results correspond to the cc-pVTZ basis \cite{dunning_basis}.}
  \label{tab_cyclobut_ens}
\end{table}

  The automerization potential energy surfaces of the four lowest lying states (S$_0$, T$_1$, S$_1$, and S$_2$) are depicted in Figure \ref{fig_cyclobut}. The automerization path corresponds to the linear interpolation between the S$_0$ optimized structure of cyclobutadiene and the S$_0$ optimized structure of its transition state. Figure \ref{subfig_cyclobut_14} contains the results of TCCSD and MR TCCSD methods with CAS(20,14), Figure \ref{subfig_cyclobut_19} results with CAS(20,19). 
  Numerical values of the S$_0$ automerization energy barriers for all the employed methods plus the MR AQCC results of Eckert-Maksic \textit{et al.} \cite{em_2006} are collected in Table \ref{tab_cyclobut_barriers}.
  The adiabatic (i.e. with respect to the optimized geometries of excited states taken from Ref.~\citenum{em_2006}) energy gaps between the S$_0$ transition state and the lowest lying excited states (T$_1$, S$_1$, and S$_2$) are presented in Table \ref{tab_cyclobut_ens}.

  Since the description of the ground state automerization is a CAS(2,2) problem (two degenerate orbitals for the transition state), we expect MR CCSD method to work well. On the other hand, single reference CCSD fails in the description by dramatically overshooting the energy barrier, which can be seen in Table \ref{tab_cyclobut_barriers}.
  In Figure \ref{fig_cyclobut}, one can observe that MR CCSD and MR TCCSD potential energy curves are very parallel, with MR TCCSD ones shifted towards lower energies. This energy decrease is what is expected (and desired), since the exact (within the given DMRG space) single and double CC amplitudes are employed for the DMRG active spaces in case of the MR TCCSD method. It is also more pronounced for the larger 19-orbital DMRG active space (see Figure \ref{subfig_cyclobut_19}).
  The TCCSD S$_0$ potential energy curves copy the MR TCCSD counterparts in the $\lambda$ region from 0 to 0.6. They, however, differ for the transition state ($\lambda = 1.0$) and also for $\lambda = 0.8$, where the bias of the results towards the choice of a reference from equally important determinants is expected for TCCSD.
  The S$_0$ TCCSD relative energies are less sensitive to the size of the DMRG active space than MR TCCSD energies. For the 19-orbital DMRG space, the difference between TCCSD and MR TCCSD energy barriers equals 1.9 kcal/mol.
   Our best result [MR TCCSD(20,19)] on the S$_0$ energy barrier corresponds to 9.13 kcal/mol and is slightly higher than the best MR AQCC result of Eckert-Maksic \textit{et al.} \cite{em_2006} (8.8 kcal/mol). When adding the MR AQCC ZPV correction (2.5 kcal/mol \cite{em_2006}), we end up with 6.63 kcal/mol, i.e somewhere in the middle of the experimental range (1.6 - 12 kcal/mol \cite{carpenter_1982}).
  Interestingly, the differences between the T$_1$ TCCSD ($M_S = 1$ component) and MR TCCSD ($M_S = 0$ component) energies are more pronounced for the larger DMRG active space (roughly 1 kcal/mol). 
  The situation is, however, quite different for the adiabatic excitation energies (Table \ref{tab_cyclobut_ens}). 
  Correct description of the excited states requires larger than CAS(2,2) space. As can be seen in the Table \ref{tab_cyclobut_ens}, 
  MR CCSD gives 
  roughly 17 kcal/mol difference when compared to the reference MR(4,4) AQCC  data in case of S$_2$ state. 
  MR TCCSD results are very consistent and the maximum difference with respect to MR AQCC is roughly 4 kcal/mol. 
  One can also observe rather significant improvement of the S$_0$ $\rightarrow$ S$_1$ MR TCCSD excitation energy when going from CAS(20,14) to CAS(20,19).

\subsection{Torsion of tetramethyleneethane}

\begin{figure}[ht]
\caption{\label{TMEscheme} Torsion reaction coordinate in the TME molecule.}
\includegraphics[scale=0.2]{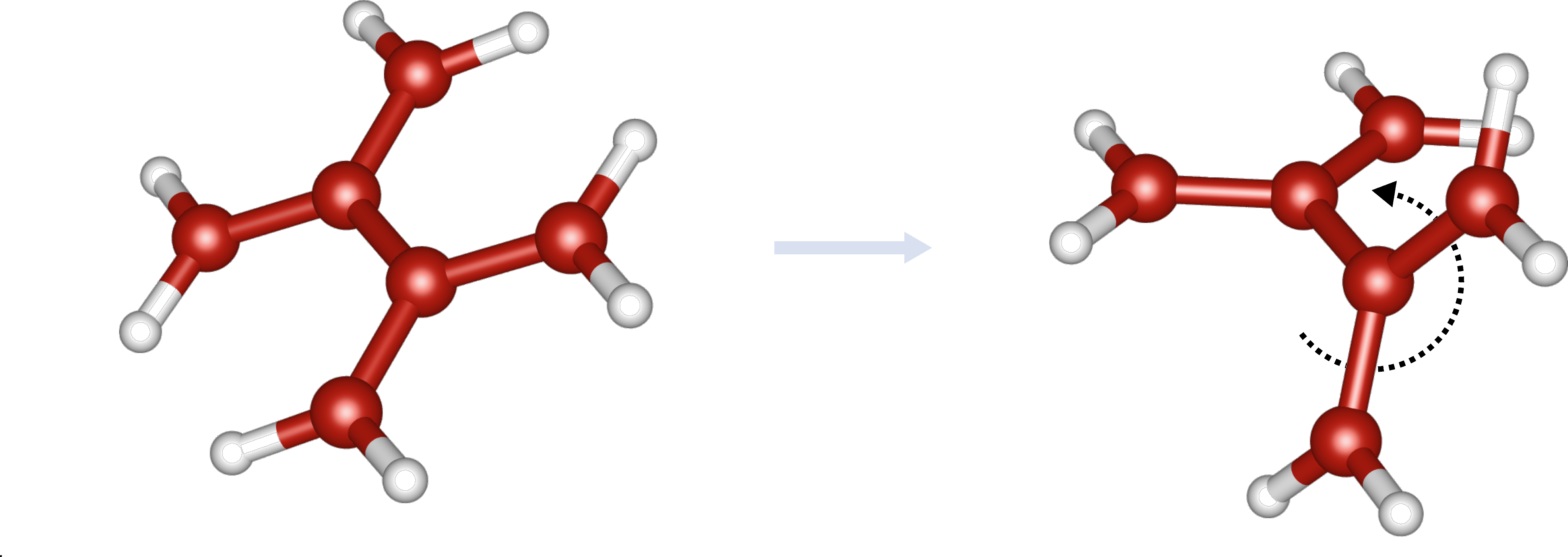}                  
\end{figure}

Tetramethyleneethane (TME) is the simplest disjoint non-Kekul\'{e} diradical, which was 
 first synthesized by Dowd \cite{dowd_1970}.
 Due to its complex electronic structure, which requires a balanced description of both static and dynamic correlation, it has been often used as a benchmark system for the state-of-the-art multireference computational methods \cite{pittner_2001, bhaskaran-nair_2011, chattopadhyay_2011, pozun_2013, demel_2015}. 
 Its complexity comes out of the fact that it contains a nearly degenerate pair of frontier orbitals, which tend to be localized on separate allyl subunits \cite{pozun_2013} and  are occupied by two electrons. 
 Moreover, TME possesses a degree of freedom corresponding to the rotation about the central C-C bond (maintaining $D_2$ symmetry, see Figure \ref{TMEscheme}) and the energetic ordering of these two frontier orbitals and consequently their occupation in the lowest singlet state changes along the rotation 
 As a result, determining the relative stability of the lowest singlet and triplet states and the rotation barrier in the singlet state turns out to be a big challenge for both experimental and theoretical methods.

\subsubsection{Computational}

As noted by Pozun et al. \cite{pozun_2013},
a relatively large basis with f functions on the C atoms is needed to reproduce the maximum on the singlet PES at 45 degrees,
cc-pVTZ being the required minimum.
In our previous study \cite{ourtmebenchmark} where we performed a FCIQMC benchmark calculation of TME, we employed the cc-pVTZ basis set and then constructed a set of 60 CASPT2(6,6) natural orbitals to which the FCIQMC calculation    was confined due to
its computational limitations.

In this work, we keep using the cc-pVTZ basis set, however, we do not employ the CASPT2 orbitals from the previous work, for several reasons. Firstly, we did not aim to repeat
a FCI-quality benchmarking which would require such a severe limitation of the virtual orbital space. Secondly, we aimed to test the multireference methods in a simpler
computational setup that better corresponds to the workflow of their intended use in applications and
which would be unnecessarily complicated by the CASPT2 step.
Finally, it turned out that in the orbitals from the previous study, which were 
computed using ORCA without a symmetry adaptation to the abelian point group, it was
rather difficult to analyze the character of excited states resulting from the DMRG calculations.
Since such an analysis is essential to determine which states to use in the multireference CC analysis for the MR tailored CC method, in this work we
computed symmetry-adapted CASSCF(6,6) orbitals using Molpro and transformed them
to a form suitable for ORCA and NWChem programs for subsequent calculations.
The DMRG calculations were carried out with Budapest QCDMRG code \cite{budapest_qcdmrg}, while TCC and MRTCC calculations were performed with our implementation in ORCA \cite{orca}. The MkCCSDT reference calculations were
performed by our local version of NWChem.
We have excluded the 1s orbital of carbon atoms from the correlation treatment.
In the MkCCSDT calculations, the highest 150 virtual orbitals have been excluded too, due to the extreme computational cost. 
As described above, we performed calculation for both canonical orbitals as obtained from the CASSCF and for a ``rotated'' set where HOMO and LUMO were rotated about $\pi/4$.

\begin{figure}[ht]
\caption{\label{TMEentropy} Entropies of the CASSCF(6,6) orbitals in the TME molecule as computed by DMRG.}
\includegraphics[scale=1.1]{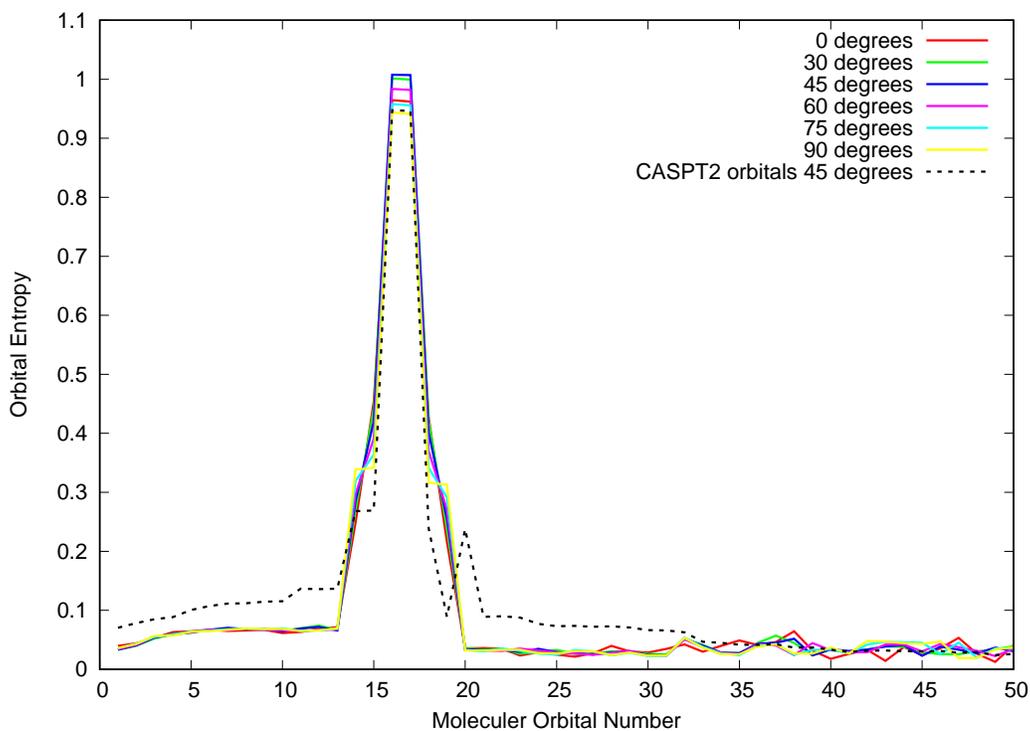}                  
\end{figure}

In order to determine suitable active spaces for the DMRG calculations,
we first computed the orbital entropies in a window of 50 orbitals using DMRG with a modest bond dimension. The results are displayed in Fig.~\ref{TMEentropy}.
It can be seen that (i) the entropies are almost independent on the torsion angle,
(ii) the two frontier orbitals with entropy close to one are form a minimal (2,2)
model space for the Hilbert-space MRCC treatment, (iii) orbitals with large entropies
form a minimum DMRG active space (6,6), while other orbitals have rather small entropy values. Selection of orbitals with entropy greater than 0.05 did yield the
DMRG active space (22,14). Note also, that the CASPT2 orbitals from the previous study had somewhat larger entropies than the presently used CASSCF ones --- cf. the dashed line
in Fig.~\ref{TMEentropy}. This can be explained by the ``compression'' of the dynamic
correlation into the CASPT2 natural orbitals.
In all production DMRG calculations (those used for generation of the active space CC amplitudes), we have employed the dynamical block state selection (DBSS) procedure \cite{legeza_2003a, legeza_2004} with the truncation error criterion set to $5 \cdot 10^{-6}$, which resulted in bond dimensions varying in the range of $1000-8000$.
In order to select excited states dominated by excitations in the MRCC model space (2,2), it was necessary to compute up to 12 excited states and to perform this selection
by hand at each value of the angle, since the order of the excited states was not constant, however, the selection could be done unambiguously.
These states were then subject to the MRCC amplitude analysis (\ref{mrccanalysis1}),
which yielded the active CC amplitudes for each reference configuration for the subsequent MRTCC treatment.
The Brillouin-Wigner MRCC and MRTCC calculations employed the a posteriori size-extensivity correction \cite{bwcc-hubaccorrection}.

\clearpage

\subsubsection{Results and Discussion}

\begin{figure}[ht]
\caption{\label{TME3x3SR} Singlet and triplet states of TME along the torsion curve computed using single reference CC and tailored CC methods.}
\begin{tabular}{ccc}
\includegraphics[scale=0.2,angle=-90]{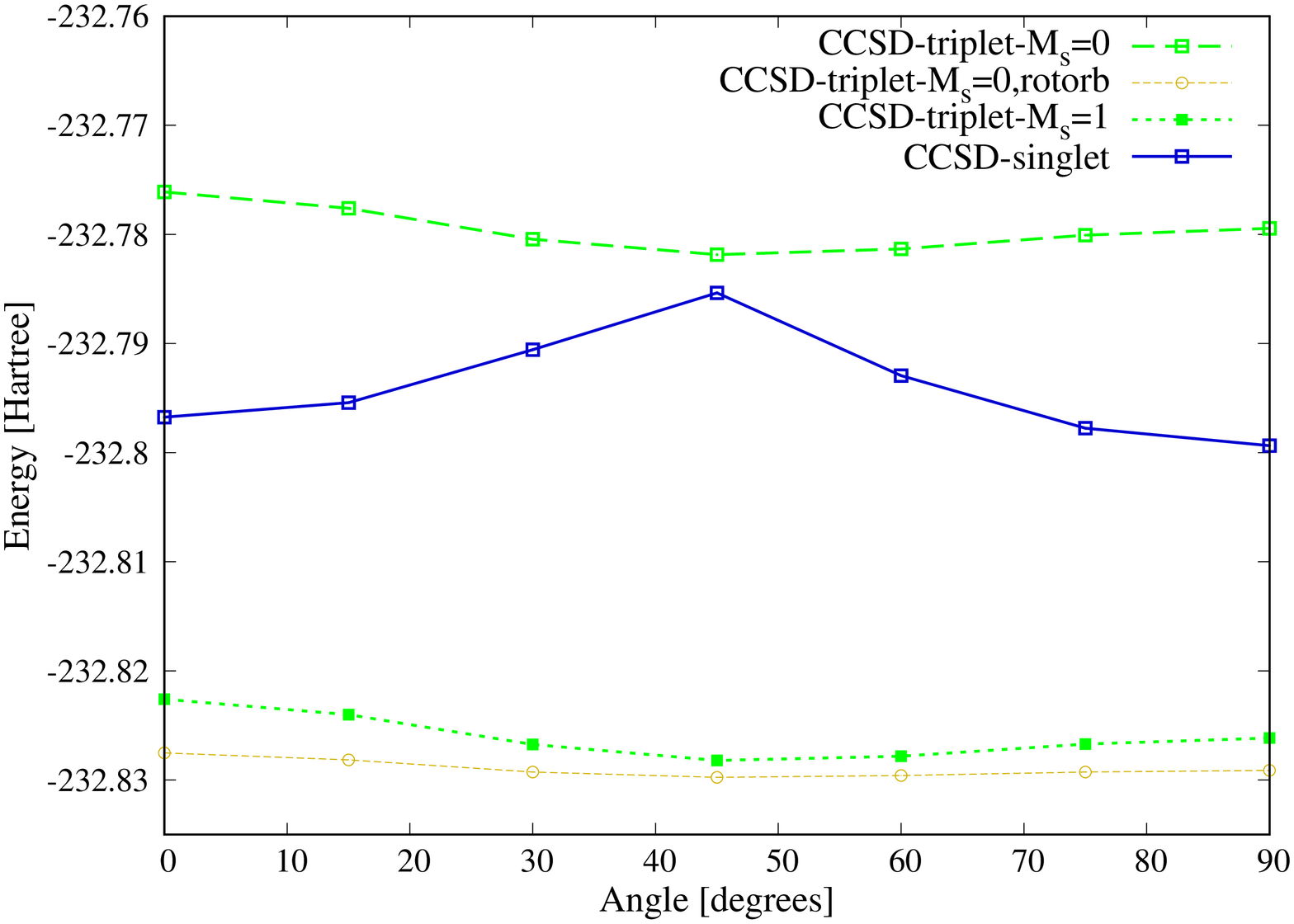}&
\includegraphics[scale=0.2,angle=-90]{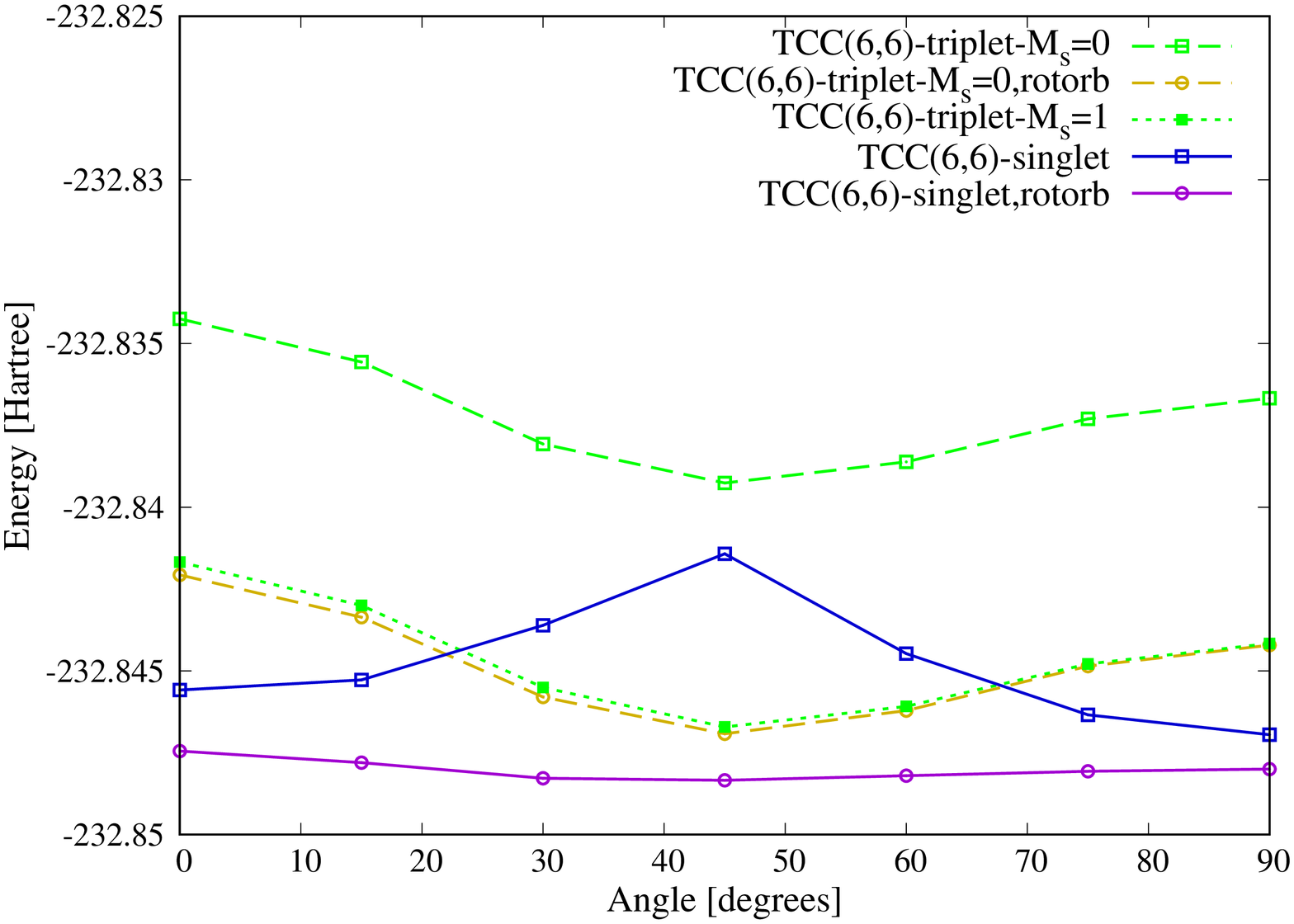}&
\includegraphics[scale=0.2,angle=-90]{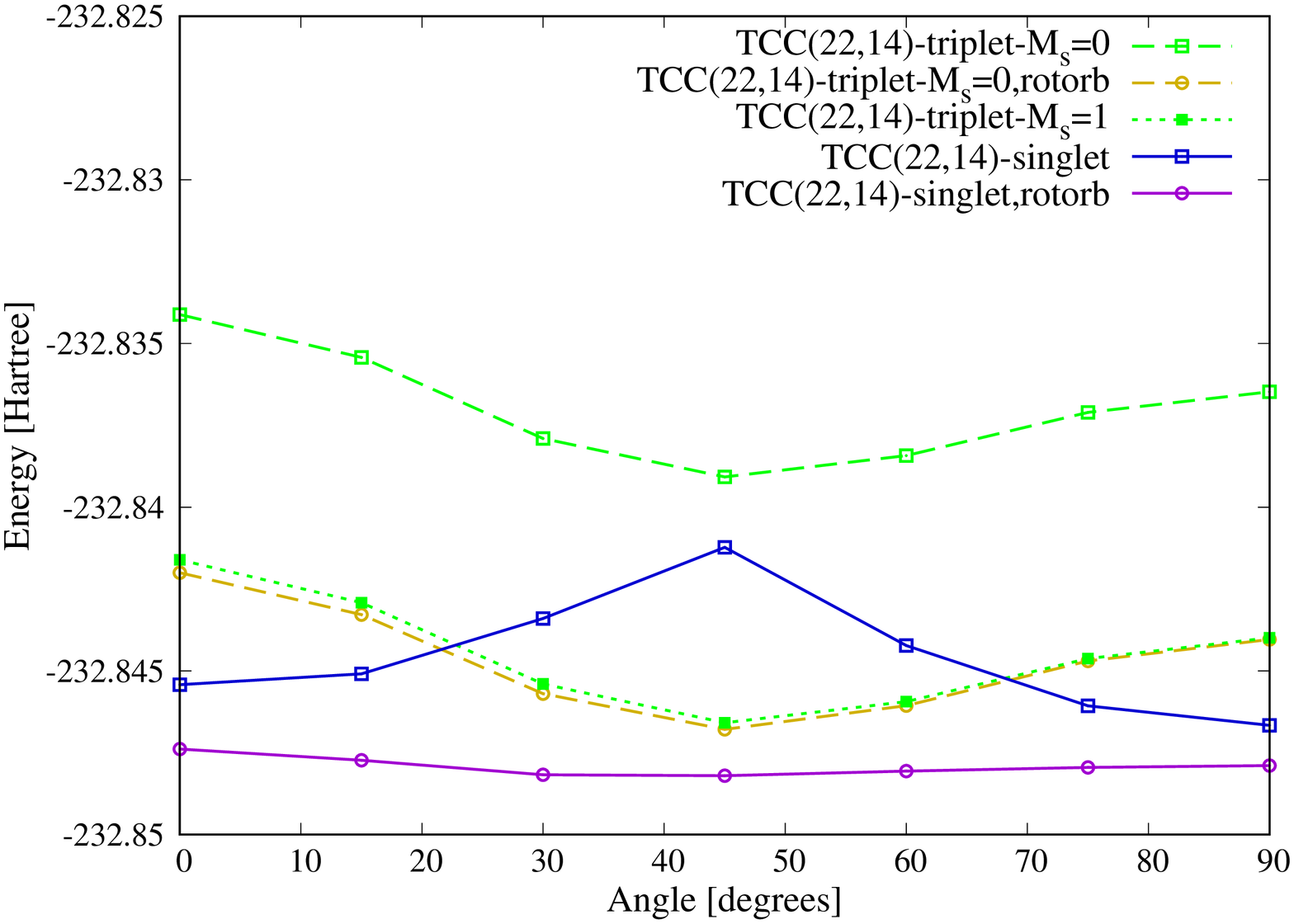}
\end{tabular}
\end{figure}

We first performed single-reference calculations employing the CASSCF(6,6) orbitals (both canonical and rotated ones). The results can be seen in Fig.~\ref{TME3x3SR} and Table~\ref{tmetable}. For the convenience of the reader, this table summarizes also the results of our previous work on TME \cite{ourtmebenchmark}. Somewhat surprisingly,
even the standard CCSD method did yield qualitative correct singlet-triplet gap as well as
the shape of the singlet state with a maximum at 45$^\circ$, provided the $M_S=0$ triplet component has been considered. The $M_S=1$ triplet curve as well as $M_S=0$ triplet in rotated orbitals are rather flat and lying lower in energy. With standard CCSD it was not possible to converge to the singlet in rotated orbitals.
The single-reference TCC method in both (6,6) and (22,14) DMRG spaces did yield
qualitatively correct ST gap both in canonical and rotated orbitals, but the singlet
barrier was correctly reproduced only in the canonical ones. This is a consistent behavior to
what we observed in the previous study\cite{ourtmebenchmark}, where CASPT2 natural orbitals were employed, but the HOMO-LUMO rotation was not investigated. However, in contrast to the CASPT2 orbitals, here the TCC results were almost independent on the size of the DMRG active space,
which is probably due to substantially lower orbital entropies of the CASSCF orbitals outside the minimal (6,6) space.

\begin{figure}[ht]
\caption{\label{TME3x3MR} Singlet and triplet states of TME along the torsion curve computed using three variants of MRCC and MRTCC methods.}
\begin{tabular}{ccc}
\includegraphics[scale=0.2,angle=-90]{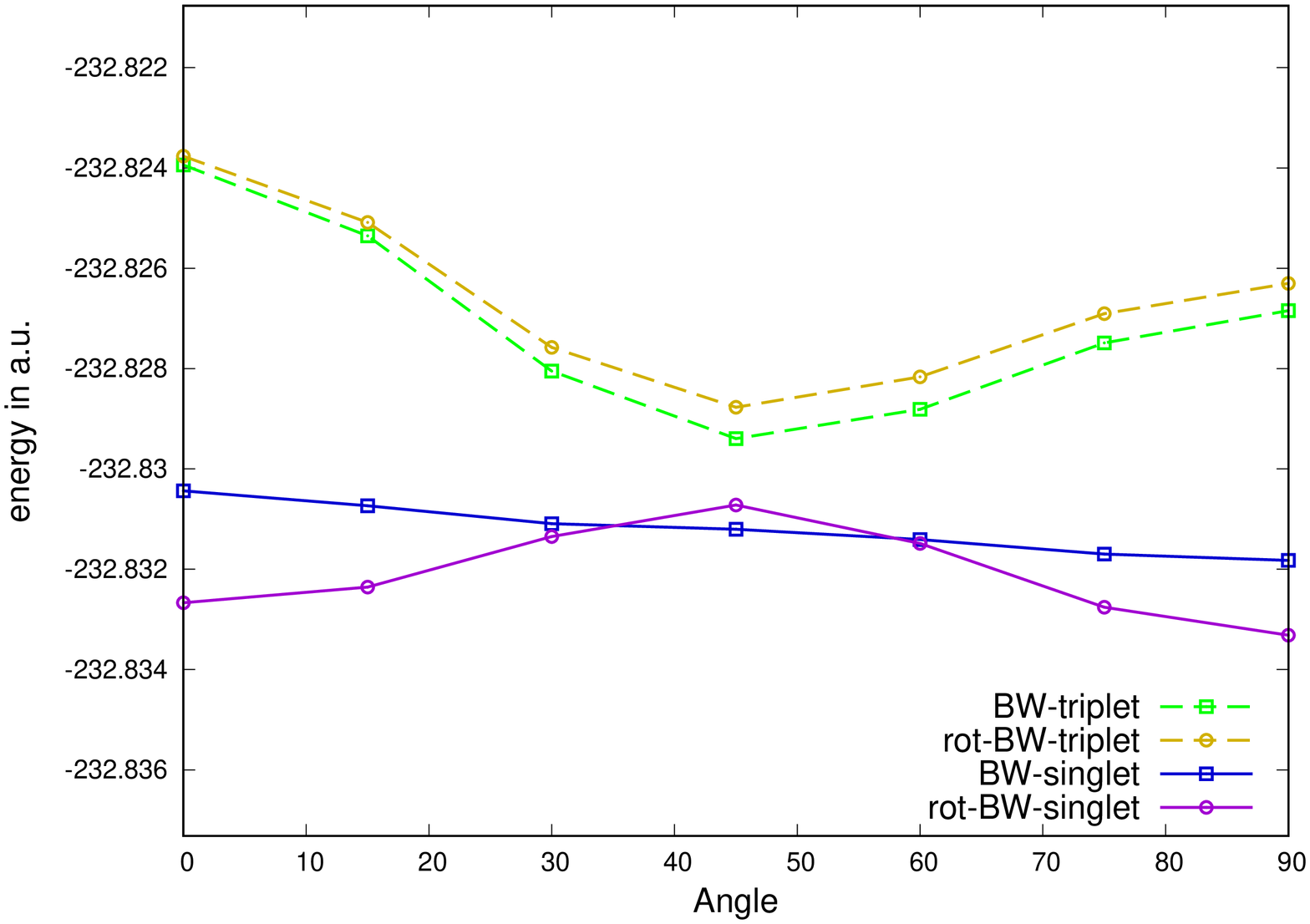}&
\includegraphics[scale=0.2,angle=-90]{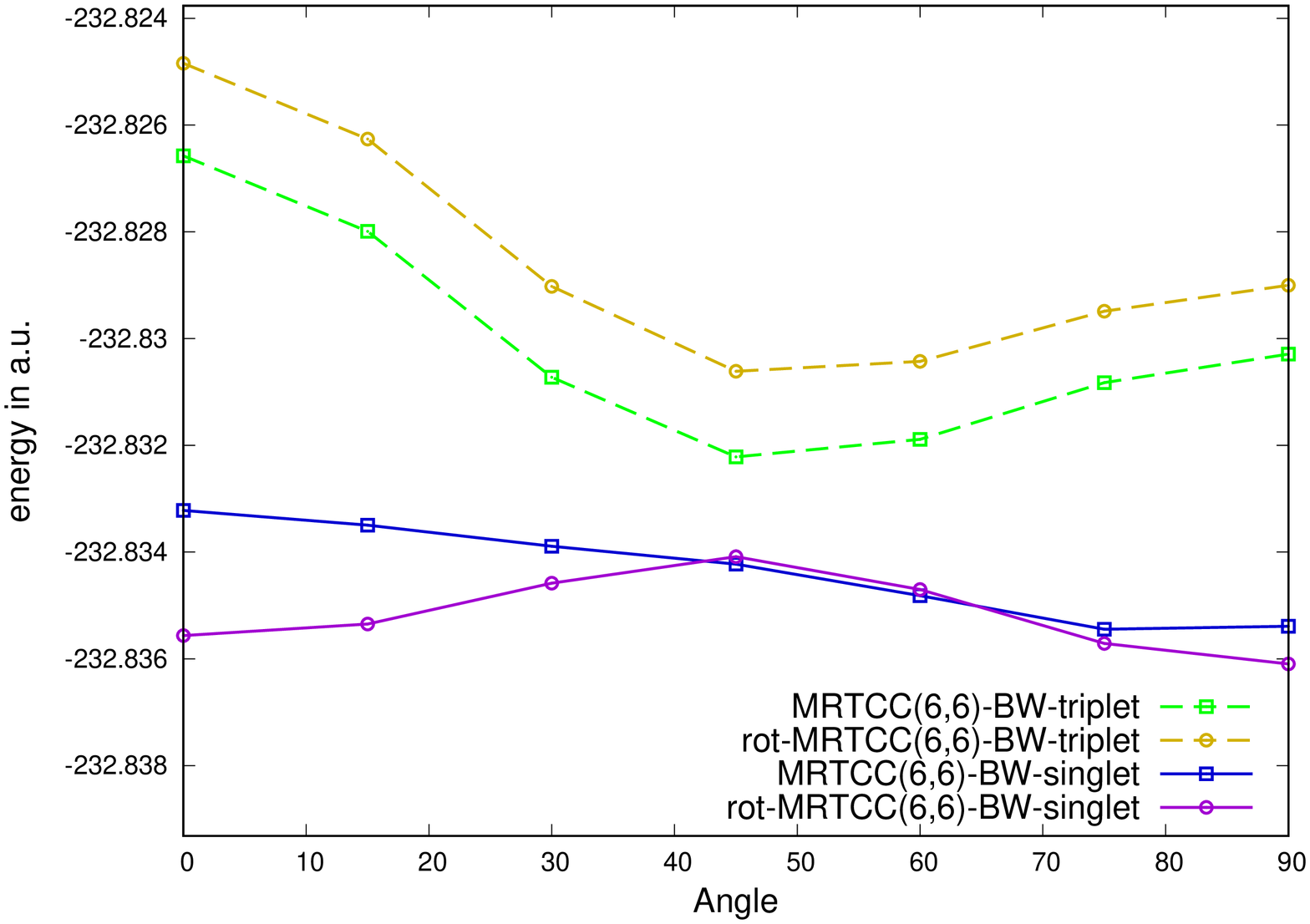}&
\includegraphics[scale=0.2,angle=-90]{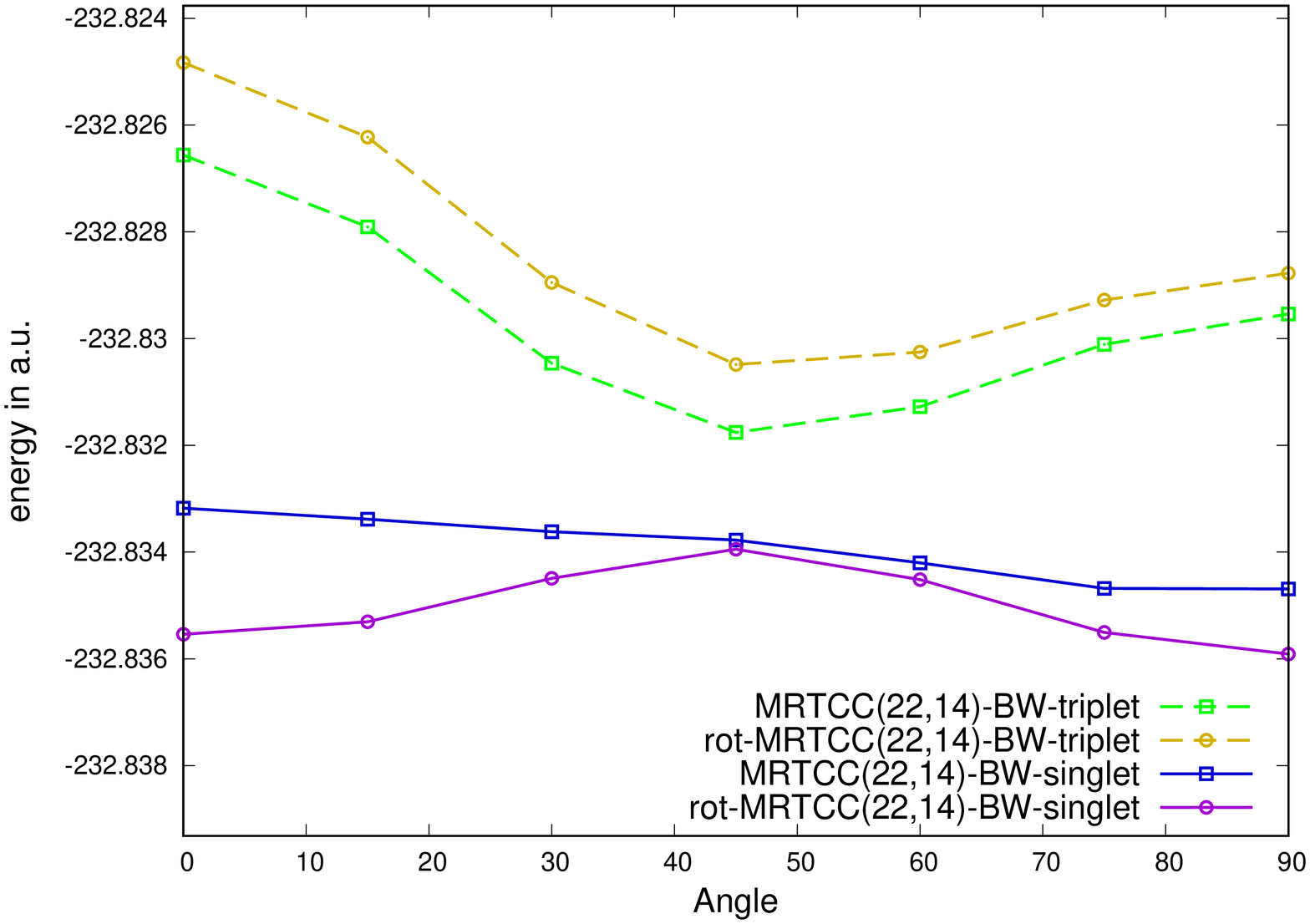}\\
\includegraphics[scale=0.2,angle=-90]{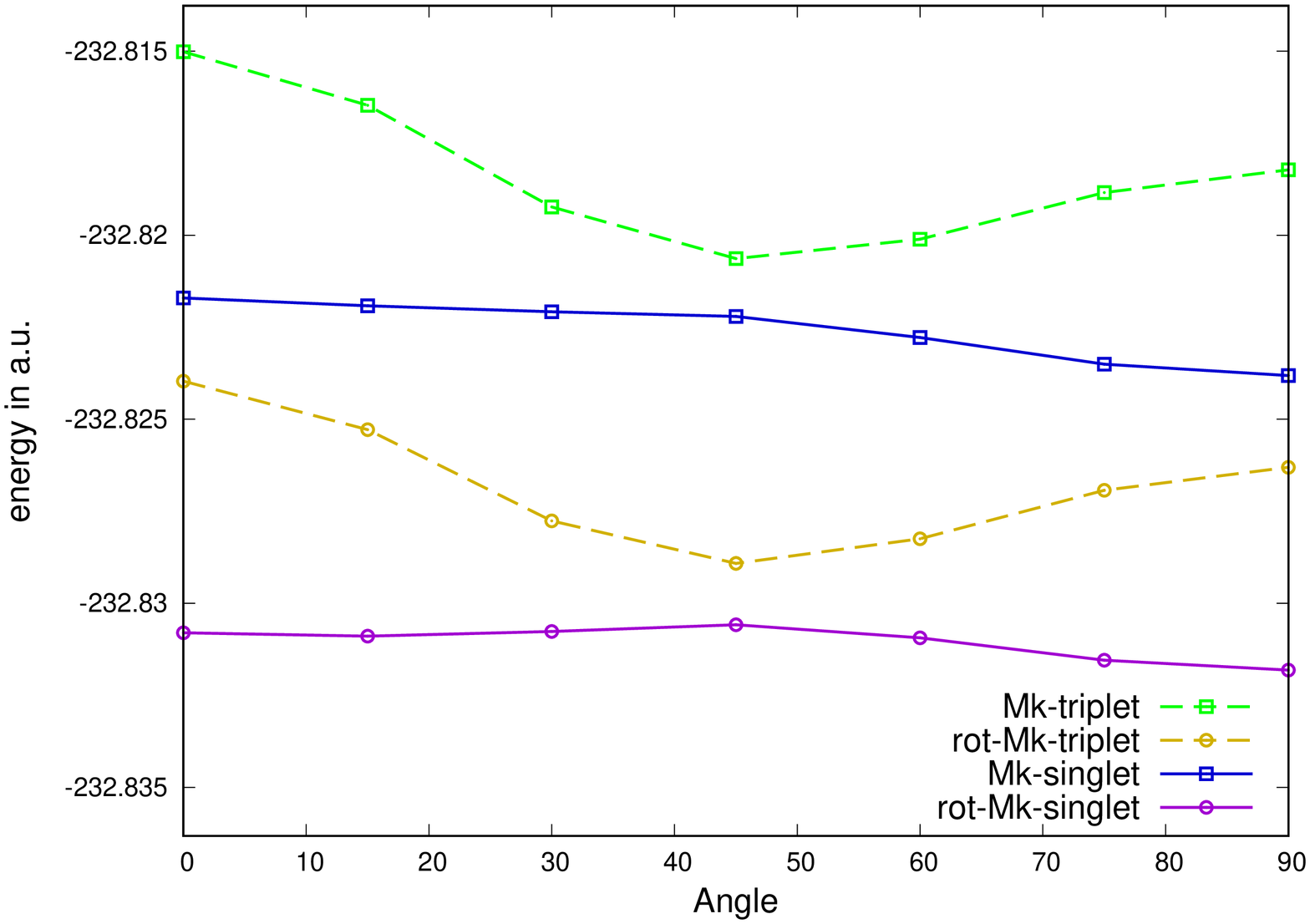}&
\includegraphics[scale=0.2,angle=-90]{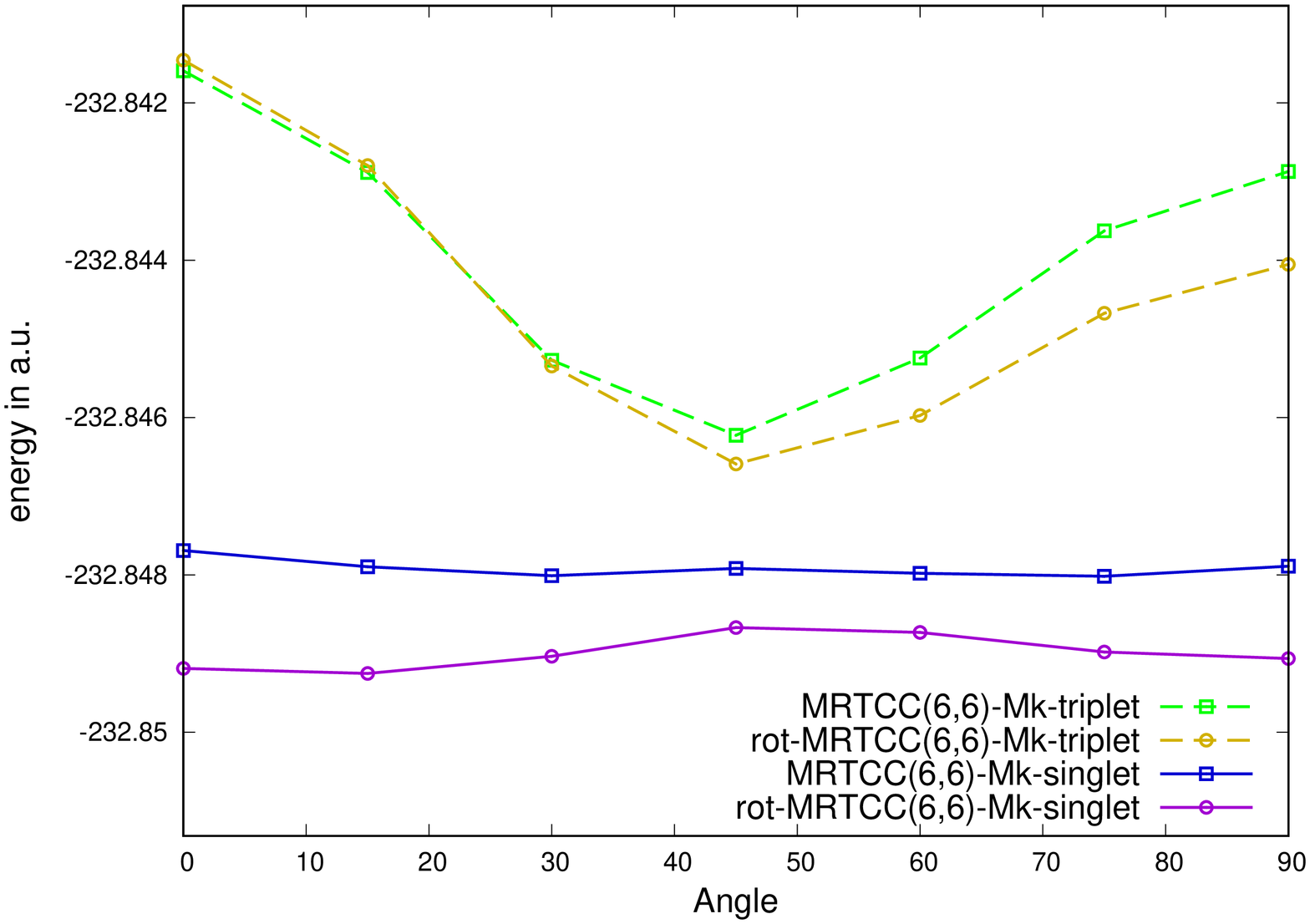}&
\includegraphics[scale=0.2,angle=-90]{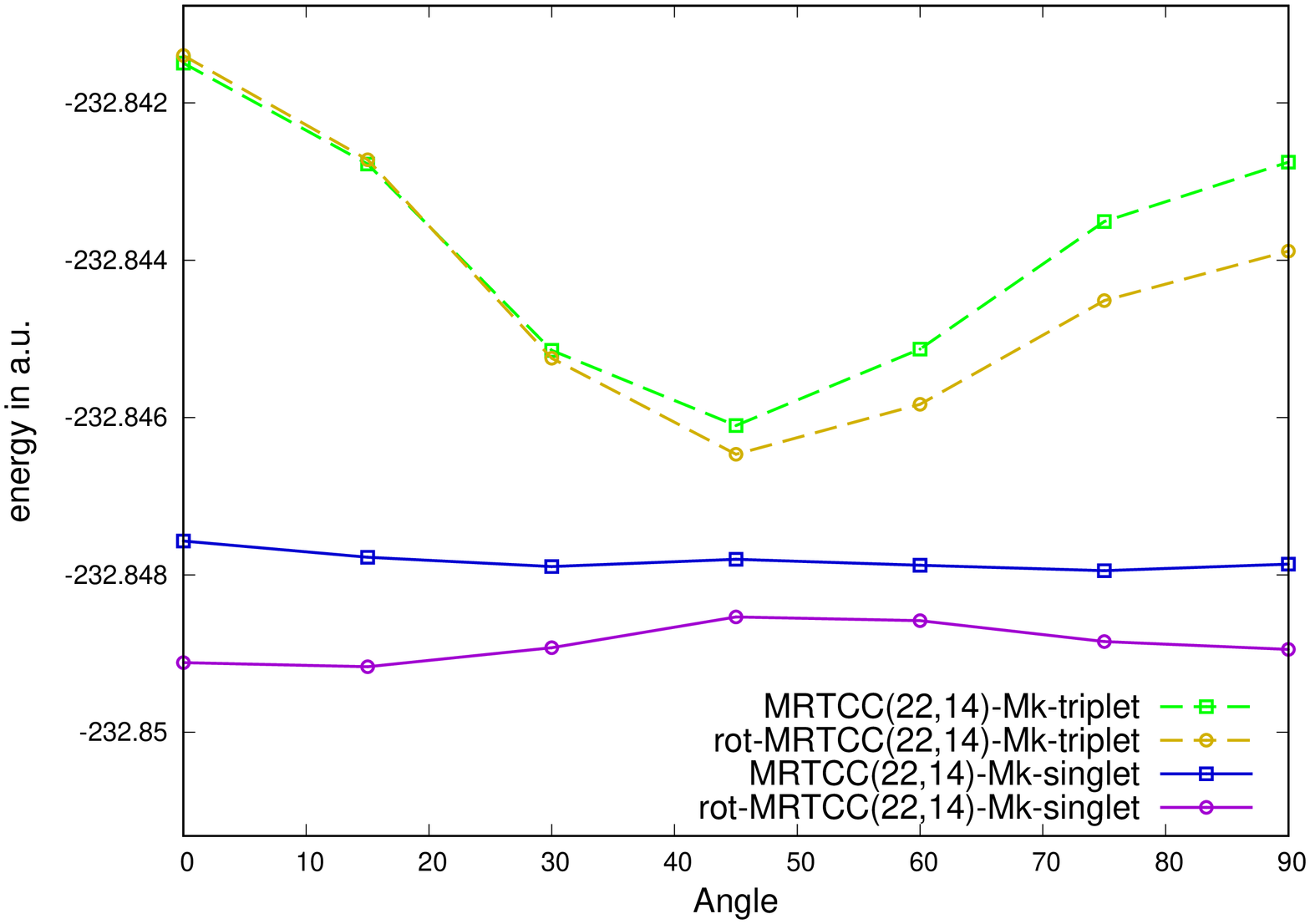}\\
\includegraphics[scale=0.2,angle=-90]{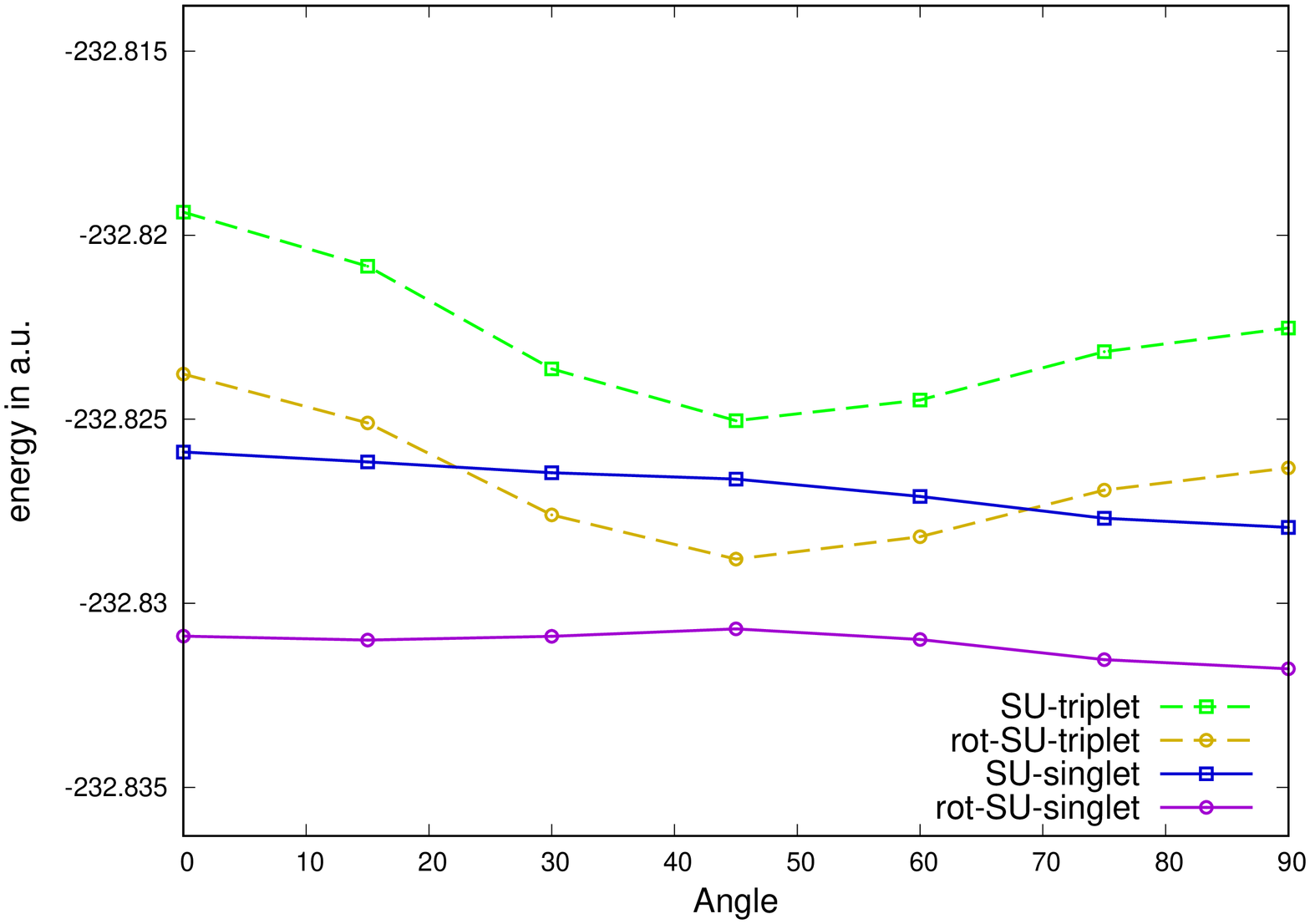}&
\includegraphics[scale=0.2,angle=-90]{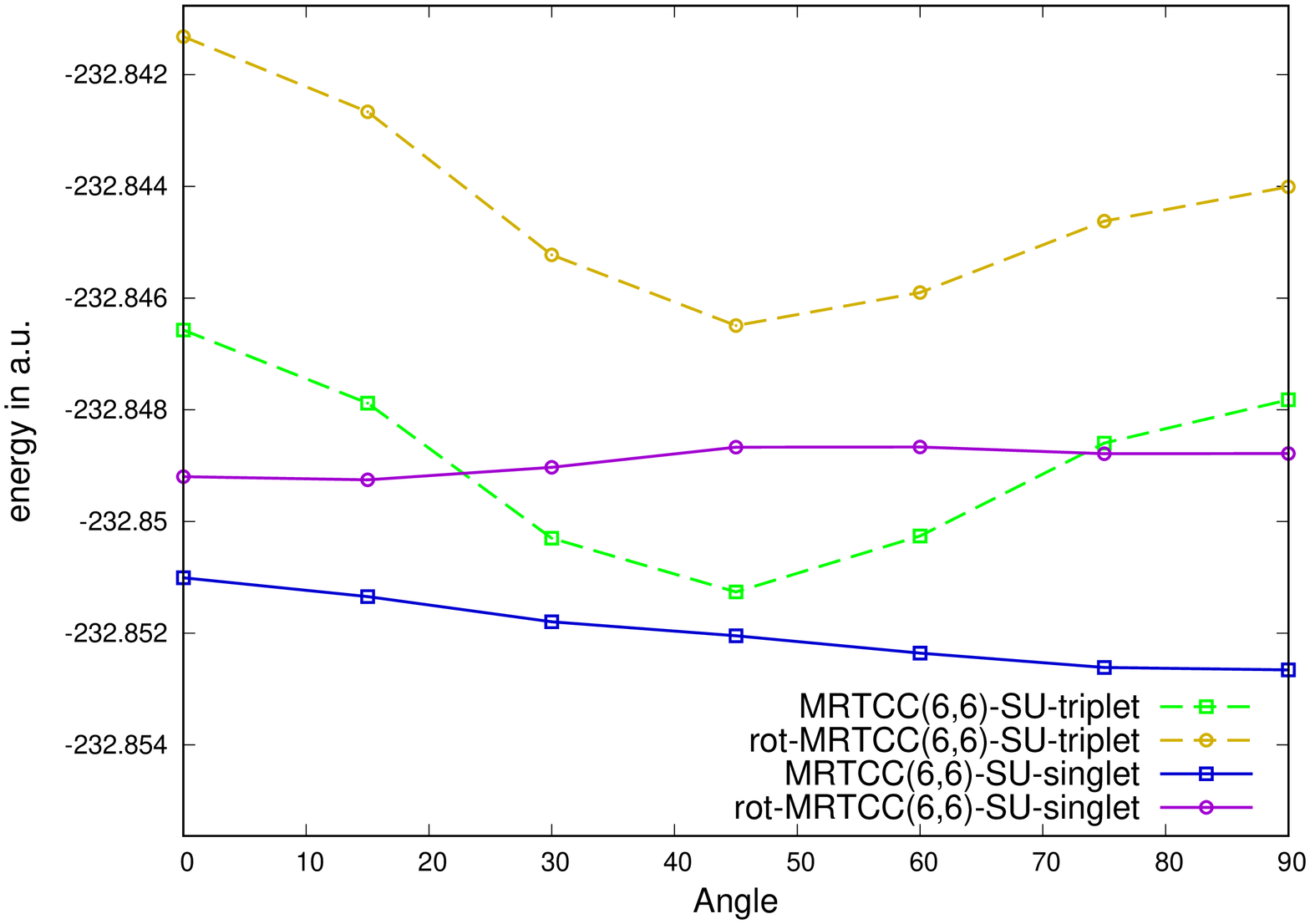}&
\includegraphics[scale=0.2,angle=-90]{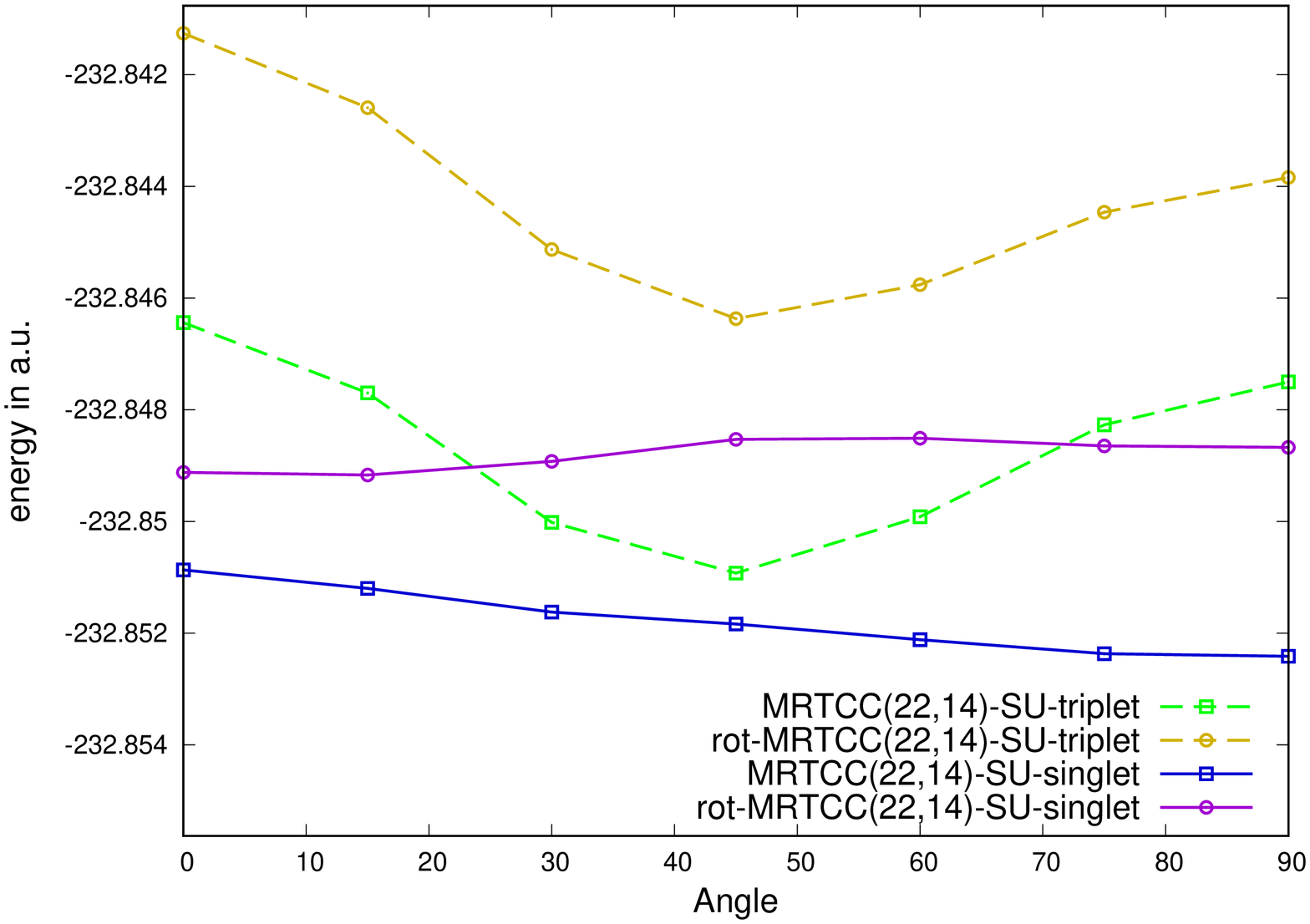}\\
\end{tabular}
\end{figure}

Next, we performed a systematic study at the multireference  CCSD level, employing three variants of the MRCC treatment (BW, Mk, SU), both canonical and rotated orbitals,
and comparing the ``plain'' MRCC method and its tailored variants with DMRG(6,6) and DMRG(22,14) spaces. The results can be seen in Fig.~\ref{TME3x3MR} and Table~\ref{tmetable}.
As can be seen, the ST gap is qualitatively correct in all cases, although the absolute energy of the curves  depends on the particular treatment, the orbital rotation having a considerable influence. However, the singlet barrier is clearly more difficult to describe. In the canonical orbitals, none of the multireference methods (at the SD level) captures the shape of the singlet curve correctly. In the rotated orbitals, BWCC yields the most prominent maximum in the middle of the curve, regardless whether it was tailored or not. MkCC and SUMRCC yield much flatter singlet PES curves even in the rotated orbitals. It is interesting that this situation is opposite to the single reference CCSD, which did yield the right shape in the canonical orbitals. In any case, it seems that at the singles and doubles level, the methods are not reliable enough to give
the right shape of the singlet independently on the orbitals employed.
Overall, for this molecule the tailored MRCC methods do not yield any big advantage over
the original MRCC. It also turns out that the error cancellation scheme for the single-reference bias
of the original TCC method works in this case exceptionally well, so that the main advantage 
of the multireference tailored treatment, i.e. being able to describe several states on the same footing, did not come to expression. However, this might be a peculiarity present only in some molecules.

\begin{figure}[ht]
\caption{\label{TMEMkCCSDT} Singlet and triplet states of TME along the torsion curve computed using the MkCCSDT method.}
\includegraphics[scale=0.35,angle=-90]{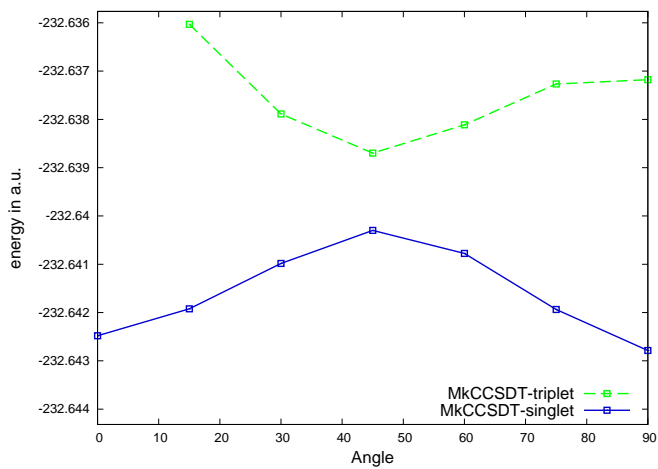}
\end{figure}

Finally, we performed a benchmark calculation employing the MkCCSDT method, cf. Fig.~\ref{TMEMkCCSDT} and Table~\ref{tmetable}. The full iterative inclusion of triexcitations
did improve the results in the canonical orbitals, yielding a well shaped maximum on the singlet curve and minimum on the triplet. The singlet barrier is very close to the FCIQMC value, while 
the ST  gaps are only a slightly higher than the DMC benchmark and the experimental value.

\begin{table}[ht]
\caption{\label{tmetable}Singlet-triplet gaps ($E_T-E_S$) and rotation barrier in the singlet state ($E_S(45^\circ)-E_S(0^\circ)$)
in the TME molecule obtained by single- and multireference methods in the cc-pVTZ basis set. In single-reference methods the $M_S=0$ triplet component has been used. See text for details on the employed orbitals. All values are in mE$_h$.}
\small
\begin{tabular}{|c|ccc|ccc|}
\hline
Method & \multicolumn{3}{|c}{Canonical orbitals} & \multicolumn{3}{c|}{Rotated orbitals} \\ \hline
& singlet barrier & $E_{ST}$ 45$^\circ$ & $E_{ST}$ 90$^\circ$ & singlet barrier & $E_{ST}$ 45$^\circ$ & $E_{ST}$ 90$^\circ$\\ 
\hline
TCCSD(6,6) &4.16 &  2.16 & 10.28 &   -0.89  & 1.42 & 3.79\\
TCCSD(22,14) & 4.19 & 2.15 & 10.18 & -0.82& 1.41 & 3.86\\
MRTCCSD(6,6)-BW	&	-1.00	&	2.01	&	5.10			&	1.48	&	3.47	&	7.10		\\
MRTCCSD(6,6)-Mk	&	-0.23	&	1.69	&	5.02			&	0.52	&	2.08	&	5.01		\\
MRTCCSD(6,6)-SU	&	-1.04	&	0.78	&	5.02			&	0.53	&	2.18	&	4.77		\\
MRTCCSD(22,14)-BW	&	-0.60	&	2.01	&	5.15			&	1.59	&	3.46	&	7.13		\\
MRTCCSD(22,14)-Mk	&	-0.23	&	1.70	&	5.11			&	0.58	&	2.07	&	5.06		\\
MRTCCSD(22,14)-SU	&	-0.97	&	0.91	&	4.92			&	0.59	&	2.16	&	4.83		\\
MRCCSD-BW	&	-0.76	&	1.81	&	4.98			&	1.95	&	1.95	&	7.02		\\
MRCCSD-Mk	&	-0.50	&	1.57	&	5.59			&	0.22	&	1.66	&	5.51		\\
MRCCSD-SU	&	-0.73	&	1.59	&	5.42			&	0.20	&	1.90	&	5.46		\\
MRCCSDT-Mk & 2.18 & 1.60 & 5.60 &&& \\

\hline
& \multicolumn{3}{|c}{60 CASPT2 orbitals} &  \multicolumn{3}{|c|}{}\\
\hline
    MkCCSD$^a$                       &  &4.4 &-1.1 &&& \\
    MkCCSDT$^a$                      &  & 2.6 & 5.5 &&&\\
    DMRG(24,25)$^a$                  &  & 1.8 & 4.8 &&&\\  
    TCCSD(6,6)$^a$&4.03 &  &  &&&\\
    TCCSD(12,12)$^a$& 2.51 &  &  &&&\\
    TCCSD(24,25)$^a$& 1.80 & 2.6 & 7.3 &&&\\
    FCIQMC$^a$       &  1.84  & 0.37 & 4.8 &&&\\
    benchmark               && 0.7$^b$ & 4.8 $\pm$ 0.5$^c$ &&&\\ \hline
\hline
\end{tabular}\\
$^a$ Our previous results from \cite{ourtmebenchmark}\\
$^b$Diffusion MC result from  \cite{pozun_2013}\\
$^c$Photo-electron spectroscopy result from \cite{clifford_1998}
\end{table}
\normalsize

\clearpage
\section{Conclusions}

We have extended the tailored CC idea to the Hilbert-space multireference CC, where a small CAS1 space employed in the MRCC contains the most strongly contributing determinants, while a relatively large CAS2 space is treated at the DMRG level.                            
The resulting DMRG-tailored MRCCSD avoids the single-reference bias of tailored CCSD, yielding more robust results in exactly or almost degenerate situations. 
For the cyclobutadiene automerization barrier, the MRTCCSD method was able to yield
better result than the single-reference TCC, where the single-reference bias leads to a too high value.
However, the improvement given by MRTCCSD with respect to MRCCSD was marginal, since the MRCC performs already good enough in this case.
Concerning the TME molecule, it turned out that here the error compensation scheme for the single-reference tailored CC works extremely well, 
so that the MR tailored methods at the SD level did not yield significantly better results of singlet-triplet energy gaps.
Moreover, it turned out that the ST energy gap is much easier to describe than the 
shape of the singlet PES. The latter is at the CCSD level strongly dependent on the underlying orbitals and on the particular version of the multireference method. 
This suggests an increase of the excitation level is needed.
The multireference CCSDT calculation did yield results in agreement with the FCIQMC benchmark and
experiment, confirming the importance of triexcitations. However, it is too expensive to be practicable and
we thus plan to include perturbative triples correction to the DMRG-tailored MRCC methods, as this
might improve the shape of the singlet potential energy curve in TME and 
better include the dynamic correlation in general.

\clearpage

\section*{Acknowledgments}

The work of the Czech team has been supported by the \textit{Czech Science Foundation} Grant \mbox{No. 18-24563S and 19-01897S}.
	We also highly appreciate the generous admission to computing
facilities owned by parties and projects contributing to the
National Grid Infrastructure MetaCentrum provided under the
program “Projects of Large Infrastructure for Research,
Development, and Innovations” (no. LM2010005) and
	by the project "e-Infrastruktura CZ" (e-INFRA CZ LM2018140 ) supported by the Ministry of Education, Youth and Sports of the Czech Republic.
    J. Brandejs was supported by the Charles University in Prague (grant no.\ CZ.02.2.69\slash0.0\slash0.0\slash19\_073\slash0016935).
\mbox{{\"O}. Legeza}  has been supported by the Hungarian National Research, Development and Innovation Office (NKFIH) through Grant
Nos.~K134983 and TKP2021-NVA-04, by the Quantum Information National Laboratory of Hungary, and by the Hans Fischer Senior Fellowship programme funded by the Technical University of Munich --
Institute for Advanced Study.
Authors also acknowledge support by the Center for Scalable and Predictive methods for Excitation and Correlated phenomena (SPEC),
funded as part of the Computational Chemical Sciences Program by the U.S. Department of Energy (DOE), Office of
Science, Office of Basic Energy Sciences, Division of Chemical Sciences, Geosciences, and Biosciences at Pacific
Northwest National Laboratory.
\mbox{M. Mihály} has been supported by the ÚNKP-19-3 New National Excellence Program of the Ministry for Innovation and Technology.

\let\OLDthebibliography\thebibliography 
\renewcommand\thebibliography[1]{
\OLDthebibliography{#1}                                                                                 \setlength{\parskip}{0pt}
\setlength{\itemsep}{0pt plus 0.3ex}
}

\bibliography{cc,cc2020,dmrg,relativity,reltcc,ors,mrtcc2022,perspective,tme_references}

\end{document}